\documentclass[11pt]{article}
\usepackage{epsfig, amsfonts, amsmath, amssymb, fullpage}
\pdfoutput=1
\usepackage{hyperref}
\usepackage{graphicx}
\usepackage{subcaption}
\usepackage{dcolumn}
\usepackage{capt-of}
\usepackage{caption}
\usepackage{multirow}
\usepackage{cleveref}

\usepackage{algorithm}
\usepackage{algpseudocode}
\usepackage{fullpage}

\DeclareMathOperator{\argmax}{argmax}

\newcommand{\real}{\mathbb{R}}

\usepackage{expl3}
\ExplSyntaxOn{}
\newcommand\latinabbrev[1]{
\peek_meaning:NTF.{
#1\@}%
{\peek_catcode:NTF a {
#1.\@}%
{#1.\@}}}
\ExplSyntaxOff{}

\title{\Large Single- and Multi-level Network Sparsification by Algebraic Distance}
\author{
	Emmanuel John\thanks{School of Computing, Clemson University, Clemson, SC \href{mailto:emmanuj@g.clemson.edu}{\tt emmanuj@g.clemson.edu}}
	\and
	Ilya Safro\thanks{School of Computing, Clemson University, Clemson, SC \href{mailto:isafro@clemsoin.edu}{\tt isafro@clemson.edu}}
}

\begin{document}

\maketitle
\begin{abstract}
Network sparsification methods play an important role in modern network analysis when fast estimation of computationally expensive properties (such as the diameter, centrality indices, and paths) is required. 
We propose a method of network sparsification that preserves a wide range of structural properties. Depending on the analysis goals, the method allows to distinguish between local and global range edges that can be filtered out during the sparsification. First we rank edges by their algebraic distances and then we sample them. We also introduce a multilevel framework for sparsification that can be used to control the sparsification process at various coarse-grained resolutions. 
Based primarily on the matrix-vector multiplications, our method is easily parallelized for different architectures.

\end{abstract}

\section{Introduction}
Networks are an abstract model of the relationships between discrete objects. Examples include networks of genes, consumers and generators in the power grid, and networks of friendships or followers in social communities. In order to study real world networks, they are often represented as graphs, where the vertices represent the objects and edges model the relationship or interaction between them. Modeling networks this way facilitates the analysis and understanding of many different structural properties of the underlying complex system. Several powerful software packages such as SNAP \cite{toolSnap}, Pajek\cite{toolPajek}, NetworkIt \cite{networkitRf}, NetworkX \cite{toolNetworkx}, and Gephi \cite{bastian2009gephi} have been developed to provide this capability. However, many complex networks are massive in size. For example, Facebook users post about 3.2 billion likes and comments each day \cite{Ahmed:netsampl}, Twitter has more than 190 million users and about 65 million tweets are posted each day \cite{wang2011understanding}, and the human gene network contain several million edges \cite{nr:bio-human-gene1}. Although, modeling and understanding these networks is very important in many application domains, the massive size of the network makes it often impractical to perform network analysis on the entire dataset.

In sparsification methods, we aim to select a representative sample of the corresponding graph such that some properties of the original graph are preserved.
In other words, central to sparsification is the idea that if an algorithm depends on or computes the properties that are preserved in the sparsified graph, we can expect that the results will be similar for the original graph \cite{Pili:samplingSurvey} while the algorithm will perform much faster on the sparsified graph.
Sampling is broadly being carried out in real world networks. Most network analytics consider just a sample in time of the networks under study which is usually the result of data collection limitations \cite{Ahmed:netsampl}. Thus, it is important to understand and develop \emph{scalable} methods for sampling massive networks.

There are several motivating examples for network sparsification. One obvious example is in the domain of visualization. It is often computationally intensive to render huge graphs on a computer screen as well it is hard to visually analyze such graphs. Sparsification helps to visualize a sample of the graph that reveals structural properties that would have been difficult to visualize and visually analyze in the original graph \cite{Maciej:topology,satuluri2011local}. The computational difficulty of visualization often arises from its objective, which requires solving a computational optimization problem \cite{hu2005efficient,auber2004tulip}.
Another broad application is the reduction in the cost of computational network analysis. In computing the betweenness centrality of every node in a massive network, for example, by prioritizing what edges should be retained and what should be removed, it is possible to improve the running time of the algorithms at a very minimal cost in optimality \cite{bader2007approximating}. Thirdly, graph sparsification can be applied to revealing hidden populations which are  hard for researchers to find by just looking at the entire population. For example, Salganik et. al showed that when trying to sample the population of injection drug dealers, it is difficult to sample directly as this population is hidden and so specialized sampling algorithms are needed \cite{salganik2004sampling}. Methods applied usually involve starting out with a sample of the desired population and using that as a seed for revealing the other members of the sample population \cite{Pili:samplingSurvey}. Existing methods include snowball sampling \cite{Handcock:snowball,goodman1961snowball} and respondent driven sampling \cite{salganik2004sampling}. In addition, in the case where there is an incomplete data, sampling can be used to estimate properties of the original graph. This is particularly useful in dynamic graphs \cite{Stutzbach:sampling}, graph streaming algorithms \cite{Ahmed:netsampl} and collective classification \cite{saha2013sparsification}.

There are several approaches to sampling a large graph while preserving the desired properties. An example involves formulating a mathematical programming problem to minimize the distance between the sparse graph and the original graph \cite{Pili:samplingSurvey}. However, such approaches are often quite complex and running them might be costlier than running the algorithm on the larger graph. Spectral approximation algorithms also exist \cite{spielman2011spectral}. However, those algorithms are not very fast as well and often infeasible for large graphs \cite{Pili:samplingSurvey} as they often involve hidden constants and require convergence in  eigen-problems. The more common approaches are (1) vertex sampling, which involves selecting a number of vertices from the original graph and retaining the vertex-induced subgraph, and (2) edge sampling, which involves the selection of edges and corresponding edge-induced subgraph. Other variations of edge and vertex  sampling have been developed (see \cite{Pili:samplingSurvey} for a full survey). In our method we focus on the edge sampling and also preserve the nodes from the original graph. In order to achieve this, we ensure that every node has at least one incident edge in the sparsified graph.

\subsection{Strength of Connectivity in Sparsification}
If the properties to be preserved are known beforehand, then, in many cases, it is possible to determine what kind of edges are important to preserve those properties and which ones are redundant. Thus, the sampling transformation can then be designed with the objective of retaining those edges. 
A general framework for sparsification involves:
(1) ranking the edges and assigning each edge an edge score; and (2) sampling edges based on their scores \cite{Pili:samplingSurvey}.
Scoring edges provides a motivation for rating the strength of connection between two vertices. In particular, this is extremely important in weighted networks, where the weights can be approximate, noisy or even completely missing. Different types of the connection strength have been proposed for scoring edges. We refer the reader to \cite{lindner2015structure} for a brief survey on the sparsification-relevant types of connectivity strength. The most relevant to our work is a cohort of spectral methods widely used in theoretical computer science to sparsify dense graphs such that some spectral properties are preserved. These are usually cut-based properties that are formulated using Cheeger inequality. For example, Spielman et al. introduced the edge effective resistance  \cite{spielman2011graph}. The effective resistance is computed using the linear system solver \cite{spielman2004nearly} which runs in $O(mlog^{15}n)$ time which can be time consuming to be feasible. Another example is the vectorized PageRank  \cite{chung2014ranking}. Various interpretations of the diffusion have been proposed and analyzed \cite{kondor2002diffusion,szlam2008regularization} for graph kernels. However, these methods usually suffer from impractical complexity.

Another relevant class of methods is based on the Jaccard index in which a similarity between two vertices is measured by computing the overlap in their neighborhoods. In \cite{satuluri2011local}, Satuluri et. al rated edges according to the local similarity $\text{sim}(i,j)=|N_i \cap N_j|/|N_i \cup N_j |$, where $N_i$ is the neighborhood of node $i$. This method was designed for clustering objectives assuming that nodes with larger shared neighborhoods are likely to belong to the same cluster. A global similarity threshold is then chosen for which edges are filtered. The authors also introduced a method for local sparsification in which they rate and filter edges per node by selecting the top $d_i^e$ edges ranked by their similarity score, where $e \in (0,1)$. Their method ensures that there is at least one edge per node after sparsification. We explore this property in our method.
This sparsification technique can be computationally expensive since it requires counting the number of triangles an edge is a part of. The authors, however, provided an approximation for computing the similarity. Based on the work of Satuluri et al. \cite{satuluri2011local}, local degree method favors the retention of high degree nodes - also known as hub nodes \cite{lindner2015structure}. As in the local similarity, for each node, they include edges to the top $d_i^e$ nodes. However, edges are sorted according to the degree of their neighbors in descending order. The main idea of this method is to keep edges in sparsified graph that leads to nodes with high degree. Additionally, vertex connectivity can be measured by the betweenness centrality, the shortest path length, the weight of substructures (such as spanning rooted forests, routes, overlapping paths that connect two vertices \cite{chebotarev2006proximity}) and algebraic distance \cite{chen2011algebraic} which we will discuss in Section \ref{algdist}.

\subsection{Our Contribution}
We introduce two methods for complex network sparsification that distinguish between strong and weak connectivity through neighborhoods of limited distance from the endpoints of edges.
In some networks (such as those that include geospatial information), these types of connections can be interpreted as long- and short-range connections while in other (such as social networks) as inner- and outer-community connections.
In both methods the sampling is based on the connectivity measured by the algebraic distance between nodes \cite{chen2011algebraic}. It generalizes the idea of methods that estimate the Jaccard coefficient for more distant neighborhoods through limited application of lazy random-walks (also known as algebraic distance \cite{chen2011algebraic}). In the first method (the single level approach) we demonstrate multiple settings of filtering local and global connections with the sampling similar to  \cite{satuluri2011local}. In the second method we propose a multilevel algorithm that combines the single level approach with the multilevel framework \cite{safro:relaxml} to sparsify graphs at different coarse-grained resolutions. We provide a robust method that can be tuned to preserve different network properties that are important in a variety of applications. The multi- and single level methods can both be used to either preserve the global structure or the local structure. We also discuss how our method can be parallelized and show the ruining time in OpenMP implementation. Evaluation of methods is demonstrated through comparison of several network properties with those measured on the original network. The proposed methods are implemented and available at \cite{emmanuj16}.

\section{Preliminaries}

We denote the graph underlying a given network by $G=(V, E, w)$, where $V$ is a set of vertices, $E$ is a set of edges, and $w:E\rightarrow \real_{\geq 0}$ is a weighting function on $E$ that represents the strength of connectivity between two vertices. The graph is undirected, containing no self-loops and multi-edges. For each node $i\in V$ we define its degree by $d_i$ and its neighbors by $N_i$. The clustering coefficient is a measure of the probability that neighbors of a node are connected to each other \cite{newmanNetworksIntro}. Consequently, it is a measure of the degree to which nodes in a network tend to cluster \cite{wang2011understanding}. The clustering coefficient of a node $i$ is defined as $c_i = \lambda_i / \tau_i$, where $\lambda_i$ is the number of triangle subgraphs $i$ participates in, and $\tau_i = d_i(d_i-1)/2$, i.e., the number of triples. The clustering coefficient of a graph $G$ is defined as 
\begin{equation}\label{eq:cc}
C_G = \frac{1}{|V'|}\sum_{i\in V'} c_i,
\end{equation}
where $V' = \{i\in V \mid d_i>1\}$. The diameter of a graph is defined as the maximum distance shortest path among all pairs of vertices in $G$ from the same connected component.
The resulting sparsified networks are compared with the original network using following properties: degree distribution, clustering coefficient, number of connected components\footnote{In many existing sparsification methods, the number of connected components is preserved ``artificially'', i.e., even if the edge is marked for deletion, it is not deleted if it increases the number of connected components. Here we do not restrict our algorithms with such requirement.}, diameter, betweenness centrality, PageRank centrality, and modularity \cite{newmanNetworksIntro}. 
We will use the Spearman Rank Correlation Coefficient ($\rho$) that is a measure of the correlation between two distributions. It is defined as $\rho = 1 - (6\sum p_i^2)/(n(n^2 - 1))$, 
where $p_i = x_i - y_i$, and $x_i$ and $y_i$ are the ranks computed from the scores $X_i$ and $Y_i$. 

\section{Algebraic distance} \label{algdist}
In order to determine the strength of connection of edges for the purpose of sparsification, we use the algebraic distance introduced in \cite{safro:relaxml,chen2011algebraic}. The algebraic distance of an edge $ij$ (denoted by $\delta_{ij}$) is interpreted as locally converged iterative process that propagates the weighted average of values from $N_i$ and $N_j$ initialized by random numbers \cite{chen2011algebraic}. This expresses the strength of connectivity between two nodes through their local neighborhoods. The process is essentially a Jacobi overrelaxation (JOR) or a lazy random walk with limited number of steps (see Algorithm \ref{alg1}). The algebraic distance was successfully used in several algebraic multigrid algorithms \cite{livne2012lean,BrandtBKL11} and in multilevel algorithms for discrete optimization on graphs (such as the minimum linear arrangement \cite{safro:relaxml}, and graph partitioning \cite{amg-sss12}) to reduce the order of interpolation that results in a sparsified coarse system.

\begin{algorithm}
\caption{Algebraic distance implementation: ComputeAlgDist}
\label{alg1}
\begin{algorithmic}[1]
\State {\bf Input:} Parameter $\alpha$ (in our experiments $\alpha=1/2$)
\State $\forall ij \in E$ $R_{ij} = 0$
\For{$r=0,1,2,...$} \Comment{the number of test vectors $r$ is small}
		\State $\forall i\in V$ $x_i^{(0)} \gets rand(-0.5, 0.5)$
	\For{$k =0,1,2,...$} \Comment{the number of JOR iterations $k$ is small}
			\State $\forall i\in V$ $x_i^{(k)} \gets \alpha x_i^{(k-1)} + (1 - \alpha) \dfrac{\sum_{j \in N_i} w_{ij} x_j^{(k-1)} }{\sum_{j \in N_i} w_{ij}} $
	\EndFor
	\State Rescale $x$ back to $(-0.5, 0.5)$
	\State $\forall ij \in E$ $R_{ij} = R_{ij} + (x_i - x_j)^2$
\EndFor
\State \Return $\forall ij\in E$ $\delta_{ij} \gets \dfrac{1}{\sqrt{R_{ij}+\epsilon} } $ \Comment{$\epsilon$ is sufficiently small}
\State  $\forall ij\in E$ $\delta_{ij} \gets \dfrac{\delta_{ij}}{\sqrt{d_i * d_j}}$
\Comment{optional normalization}
\end{algorithmic}
\end{algorithm}
Other stationary iterative relaxations can also be applied in a similar setting but since JOR is implicitly parallelizable using matrix-vector multiplications, we prefer to use it instead of other relaxations (such as Gauss-Seidel) that converge faster. 
Optionally, the algebraic distance can also be normalized by the square-root of the product of the weighted degrees of the two nodes to reduce extremely high strength of connection between hub nodes.

The algebraic distance will serve as the main criterion for choosing edges for sparsification in the algorithms below. Because it helps to distinguish between so called short- and long-range connections \cite{chen2011algebraic}, we will use it to demonstrate different types of sparsification in which local and global properties are preserved correspondingly to the types of algebraic distances that we choose. The short-range connections (large values of $\delta_{ij}$) will be called $\delta$-strong. The long-range connections (small values of $\delta_{ij}$) will be called $\delta$-weak. 

\section{Single-level sparsification} \label{singleLevelSection}
\begin{figure}[t]
	
        \centering
        \begin{subfigure}[b]{0.15\textheight}
                \centering
                \includegraphics[width=\textwidth]{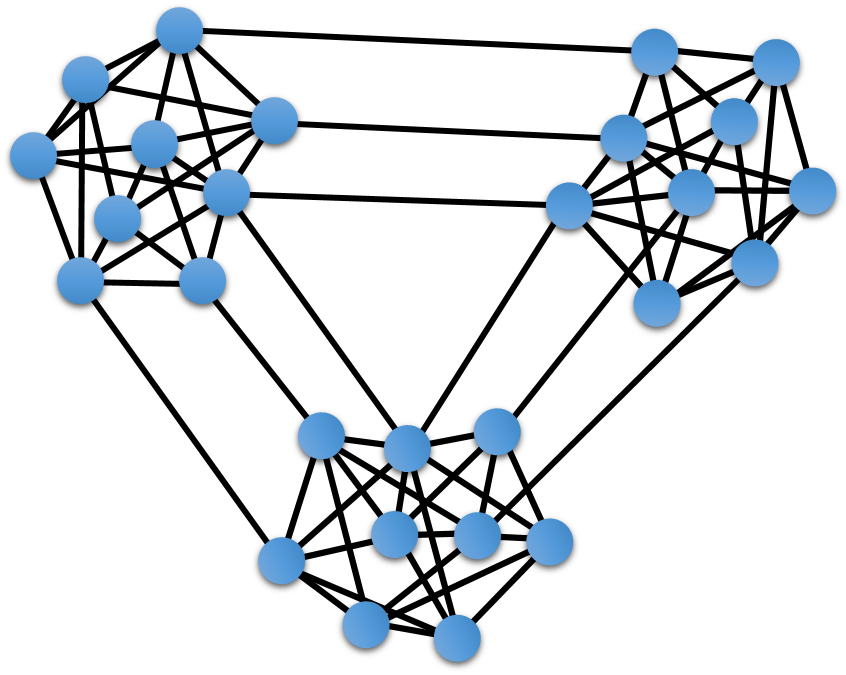}
                \caption{\footnotesize Original network}
                \label{fig:power}
        \end{subfigure}%
        \hfill
        \centering
        \begin{subfigure}[b]{0.15\textheight}
                \centering
                \includegraphics[width=\textwidth]{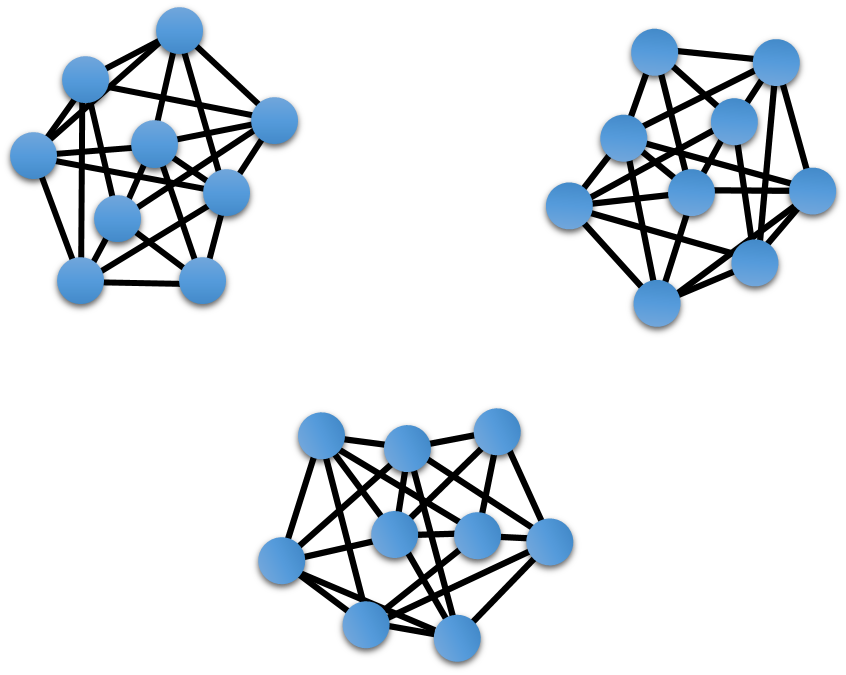}
                \caption{\footnotesize Sparsified network by eliminating $\delta$-weak connections}
                \label{fig:skg}
        \end{subfigure}%
        \hfill
        \centering
        \begin{subfigure}[b]{0.15\textheight}
                \centering
                \includegraphics[width=\textwidth]{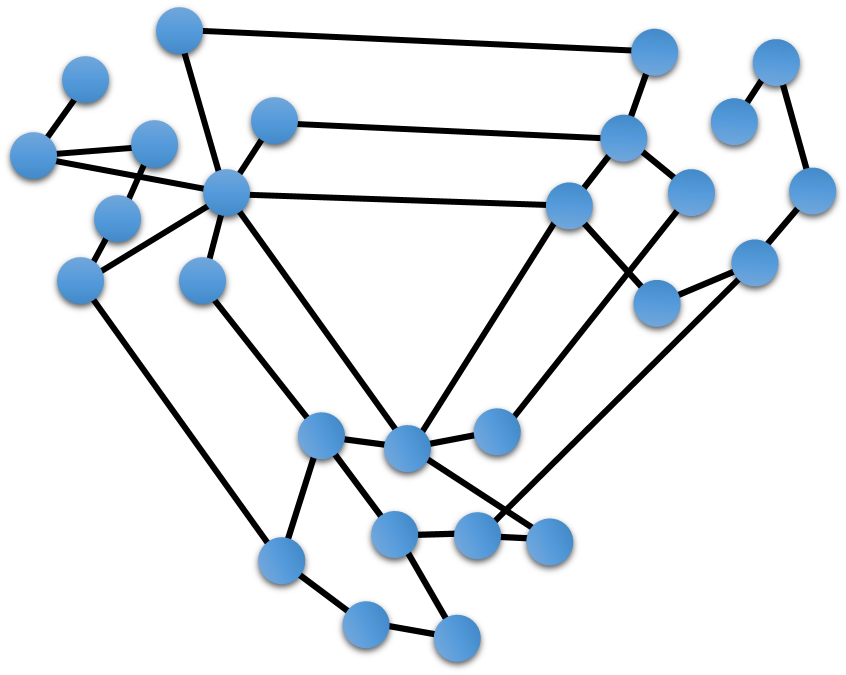}
                \caption{\footnotesize Sparsified network by eliminating $\delta$-strong connections}
                \label{fig:musk}
        \end{subfigure}%
\caption{\footnotesize An example of a small network with 3 dense clusters and sparse cuts between them (a). Sparsification of $\delta$-weak connections will result in network presented in (b). Sparsification of $\delta$-strong connections is presented in (c).}\label{fig:badgeneration}
\end{figure}

In the single-level approach we demonstrate three types of sparsification in which we filter $\delta$-weak, $\delta$-strong edges and their mixture. In all of these cases, first, for each edge in the graph, we compute the algebraic distance. Then, for each node $i$, we sample the top $d_i^e$ neighbors ranked by their algebraic distances, where $e\in [0,1]$. 
In this approach it is possible to sample for local or global structure preservation or a combination of both. 
To preserve the global structure, we select $d_i^e$ weakest connections and add them to the sparse graph (see Figure \ref{fig:badgeneration}c). Similarly, $d_i^e$ strongest connections are preserved to emphasize the importance of a local structure in the sparse graphs (see Figure \ref{fig:badgeneration}b).
\begin{algorithm}[H]
\caption{Single-level sparsification: Sparsify($G$)}
\label{alg3}
\begin{algorithmic}[1]	
\State {\bf Input:} Sparsification parameter $e$, Graph $G$
\State {\bf Output:} Sparsified graph $G_{sparse}$
\State $G_{sparse} \gets \text{ empty graph}$
\State \Call{ComputeAlgebraicDistances}{G}
\For{$i \in V$}
	\State Sort $N_i$ by $\delta_{ij}$ in ascending (or descending) order
	\State Add top $d_i^e$ edges to $G_{sparse}$
\EndFor
\State \Return $G_{sparse}$
\end{algorithmic}
\end{algorithm}

It is also possible to partially preserve both global and local structures with a slight change in the algorithm, namely, by distributing the algebraic distances into bins, and sampling the edges from all bins. In order, to distribute the algebraic distances into bins, we define the bin width $h = \frac{3.5 * \sigma}{\sqrt[3]{d_i}}$, and the number of bins $k =(\max\delta_{ij} - \min \delta_{ij})/h$, where $\sigma$ is the standard deviation of algebraic distances.
%
\begin{algorithm}
	\caption{Single-level sparsification with binning: SparsifyB($G$)}
	\label{alg4}
	\begin{algorithmic}[1]	
		\State {\bf Input:} Sparsification parameter $e$, Graph $G$
		\State {\bf Output:} Sparsified graph $G_{sparse}$
		\State $G_{sparse} \gets $ empty graph
		\State \Call{ComputeAlgebraicDistances}{G}
		\For{$i \in V$}
		\State Distribute $N_i$ into bins, each bin corresponds to edges ...
		\State Randomly select bins and edges up to $d_i^e$ to $G_{sparse}$
		\EndFor
		\State \Return $G_{sparse}$
	\end{algorithmic}
\end{algorithm}

%
\section{Multilevel sparsification}
The multilevel approach \cite{vlsicad,Walshaw2004} can be applied as a general framework for many different numerical methods. Most real-world instances are not completely random, i.e., a particular similarity or dependence between variables exists and, thus, can partially be detected to reduce their number in complex computations. Here we introduce and advocate the use of multilevel approach as a general purpose framework for network sparsification. In the heart of the proposed method lies an idea to sparsify the network at multiple scales of coarseness which, in contrast to most existing sparsification methods that sample single edges, will allow to sample clusters of edges of different sizes and $\delta$-weakness. 

It is known that the topology of many complex networks is hierarchical (or multiscale) and, thus, often might be self-dissimilar across scales \cite{Itzkovitz05, Wolpert07, Binder08, Mones12}. In such hierarchical representations, groups of nodes are aggregated into communities, which automatically bundles edges into coarse connections. Bundling the edges at different scales of coarseness will introduce different levels of $\delta$-weakness for such coarse connections which may or may not be required to be sparsified for the required analysis. For example, in the analysis of a social network, we may want to visualize only a certain type of edges that connect dense communities of small sizes, while connections between large communities and local inner connections are out of the scope. In the proposed framework, this can be achieved by creating a hierarchy of coarse representations, and sparsifying at those levels that do not correspond to the desired communities. To create a multilevel framework we use the algebraic multigrid (AMG) aggregation strategy that was introduced in \cite{SafroRB06}. For simplicity, we do not split fine nodes across the aggregates (like in some optimization problems \cite{SafroRB06,amg-sss12}) but instead cover the graph with star-like structures and coarsen them. For the completeness of paper we briefly repeat the main components of the coarsening algorithm.

Given an original graph $G$, in the multilevel framework we recursively construct a hierarchy of decreasing size coarse graphs $G_0=G, G_1,..., G_l$. The original graph is gradually coarsened into the smaller graphs until the small enough graph $G_l$ is reached. The sparsification algorithm is then run on the coarsest level and the results (i.e., edges to eliminate for sparsification) are inherited by the finer graph and the uncoarsening continues until $G_0$ is reached. In most cases, our discussion is focused on  fine-to-coarse and coarse-to-fine transformations of graphs and solutions, respectively. For this purpose, we denote the fine and coarse level graphs by $G_f = (V_f, E_f)$, and $G_c = (V_c, E_c)$, respectively.
At each level, after sparsifying edges inherited from $G_c$, Algorithm \ref{alg1} is applied to recompute algebraic distance on $G_f$. 

\begin{figure}[t]
	\centering
	\includegraphics[width=0.8\linewidth]{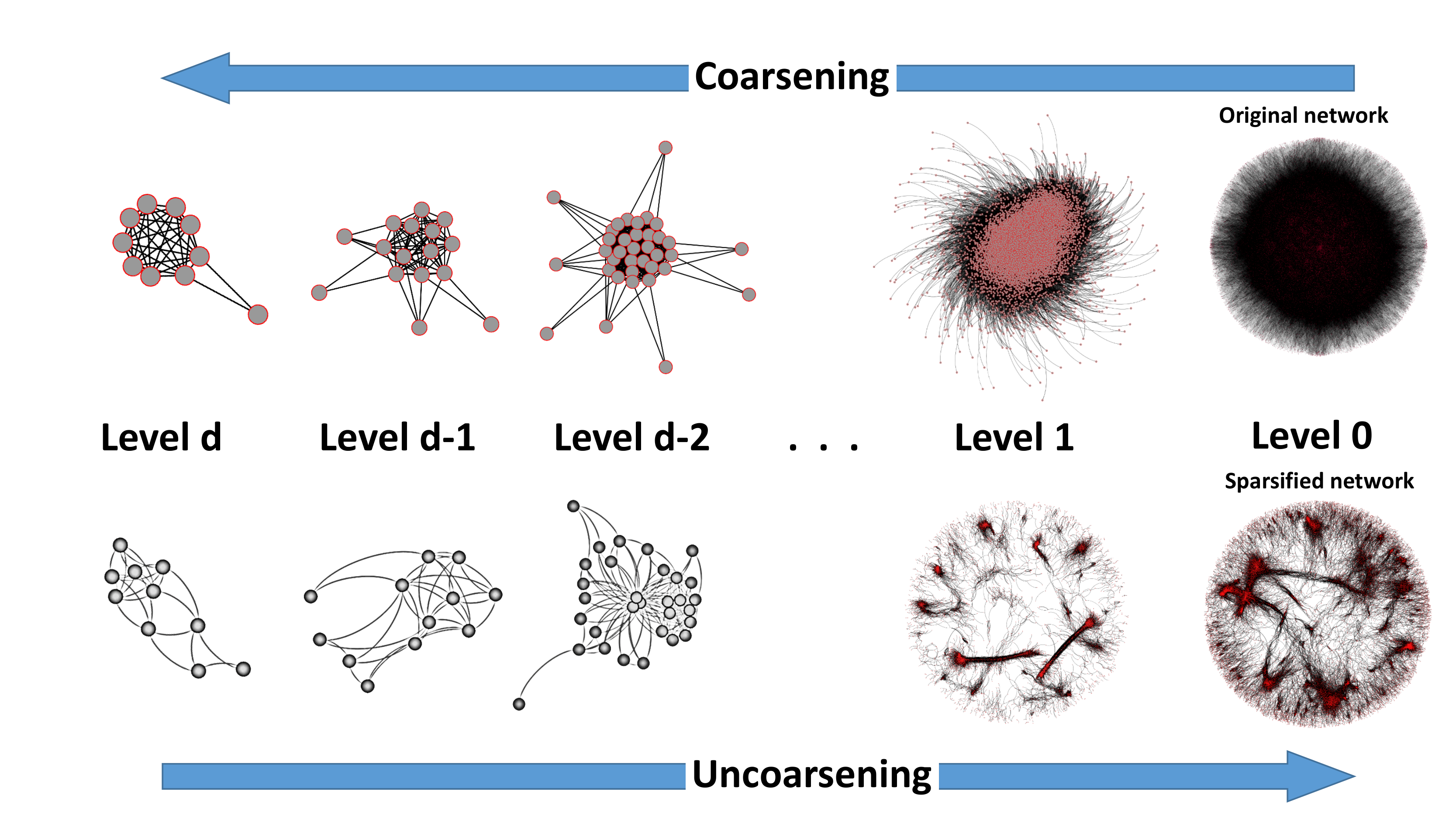}
	\caption[full-vcyle]{Complete Sparsification V-Cycle}\label{fig:cycle}
\end{figure}

\paragraph{The Coarsening} We begin with selecting a dominating set of seed nodes $C\subset V_f$ that will serve as centers of future coarse nodes in $V_c$. Setting initially $F=V_f$ and $C=\emptyset$, the selection  is done by traversal of $F$ and moving to $C$ such nodes that are not strongly coupled to those that are already in $C$. At each step $F\cup C = V_f$ is preserved, and at the end the size of $V_c$ is known, namely, $|V_c| = |C|$. After $C$ is selected, nodes in $F = V \setminus C$ are distributed to their aggregates according to the restriction operator $P \in \{0,1\}^{|V_f|\times |C|}$, where 

\begin{equation}\label{eq:P}
	P_{iJ} =
  \begin{cases}
    1    & \quad \text{if } i\in F,~J = I_c \bigg( \argmax_{j\in C} \dfrac{\delta_{ij}}{\sum\limits_{k \in C} \delta_{ik}}\bigg) \\
    1    & \quad \text{if } i\in C,~J=I_c(i)\\
    0    & \quad \text{otherwise},\\
  \end{cases}
\end{equation}

and $I_c(j)$ returns an index of coarse node $J$ that corresponds to $j\in C$. Then, the Galerkin coarsening creates a coarse graph Laplacian $L_c = P^T L_f P$, where $L_f$ is the Laplacian of $G_f$.
\paragraph{Coarsest Level} At the coarsest level, we sparsify the edges by using the single-level Algorithm (\ref{alg4}). These edges correspond to bundles of edge chains at the fine levels that connect the most distant regions in a graph, so if the goal is to preserve the global structure, the user should avoid of sparsification at deep coarse levels.
\paragraph{Uncoarsening} We initialize the solution (sparsification) of $G_f$ by  uncoarsening the edges sparsified in $G_c$. When the order of interpolation in the multilevel algorithm equals 1 (i.e., there is only one non-zero entry per row in $P$, see Eq. \ref{eq:P}), each coarse edge $IJ\in E_c$ can bundle at most two types of edges in $E_f$, namely, at most one edge that connect two seeds $I_c^{-1}(I)$ and $I_c^{-1}(J)$, and possibly multiple edges $pq\in E_f$ such that $P_{pI_c^{-1}(I)} = 1$, and $P_{qI_c^{-1}(J)} = 1$. If $IJ$ is sparsified at the coarse level, then edges of both types are sparsified at the fine level. After initialization of the fine level, we recompute algebraic distances to update the information about connectivity in the sparsified fine graph, and, then, more edges may or may not be sparsified at the fine level depending on the parameter settings. Full multilevel cycle is presented in Algorithm \ref{alg6}. Example of full multilevel cycle on a Facebook network (see fb-uf in Table \ref{ta2}) is shown in Figure \ref{fig:cycle}.

\begin{algorithm}[t]
	\caption{Multilevel sparsification of graph: MLSparsify}
	\label{alg6}
	\begin{algorithmic}[1]
		\State {\bf Input:} fine graph $G_f = (V_f, E_f)$, vector of sparsification ratios
		\State {\bf Output:} sparse graph $G_{f}^s = (V_f, E_{f}^{s})$
		\Function{ML}{$G_f$}
			\State \Call{ComputeAlgDist}{$G_f$}
			\If{$|V_f|$ is small enough}
			    \State $E_f^s \gets \Call{SparsifyB}{G_f}$ \Comment{Sparsify coarse edges}
			\Else
				\State \Call{createSeeds}{$G_{f}$} 				\Comment{Coarsening: seeds}
				\State Compute $P$ 				\Comment{Coarsening: restriction operator}
				\State $G_c \gets (L_c = P^T L_f P)$ 				\Comment{Coarsening: coarse graph}
		
				\State $G_c^s \gets$ \Call{MLSparsify}{$G_{c}$} \Comment{Recursive call to sparsify the next coarser level}
				\State $G_f^s \gets $ \Call{Uncoarsen}{$G_c^s$} \Comment{Sparsification of edges inherited from coarse level}
				\State \Call{ComputeAlgDist}{$G_{f}^{s}$} \Comment{Algebraic distances are recomputed}
				\State $G_f^s \gets \Call{SparsifyB}{G_f^s}$ \Comment{Sparsification of current level edges}
			\EndIf
		\EndFunction
		\State \Return{$G_{f}^{s}$}
	\end{algorithmic}
\end{algorithm}


%
\section{Computational Results}
\paragraph{Implementation and Evaluation} We provide C++ implementation for both the single- and multilevel algorithms in \cite{emmanuj16}. For the comparison of original and sparsified networks, we employed methods implemented in NetworKit \cite{networkitRf}. We experimented with varying degrees of sparsification, taking values of $e$ ranging from 0.1 to 0.9 (see Section \ref{singleLevelSection}). All numerical properties for the comparison are the averages over 10 runs with different random seeds for each parameter setting. The following parameter were used in computation of algebraic distance: $R=10$, $k=40$, and $\alpha=0.5$. Their robustness is discussed in \cite{chen2011algebraic}.
In addition, for the single-level algorithm (\ref{alg4}), we provide two sets of results for each graph, namely, with and without the normalization (see last step in Algorithm \ref{alg1}) of algebraic distance. In each case we experimented with sparsification of weak edges, strong edges and mixture of both.

In the multilevel algorithm (\ref{alg6}), we experimented with sparsifying at the coarsest, middle and the finest levels. In our experiments, we split the number of levels in the multilevel algorithm into 3 equal segments, and choose a parameter, level-span which determine how many levels in each segment gets sparsified. We then sparsify one segment at a time and  observe the corresponding network properties. For example, for a graph with 6 levels, with a level-span of 2, to sparsify the coarsest levels only we use the following parameter configuration: $(0.3,0.3,-1,-1,-1,-1)$, where a setting of $-1$ indicates no sparsification occurs at this level. Similarly, the middle and finest levels can be sparsified using $(-1,-1,0.3,0.3,-1,-1)$, and $(-1,-1,-1,-1,0.3,0.3)$ configuration settings respectively. However, in our implementation, users can specify any combination of settings for different levels. The sparisification ratio (ratio of number edges in the sparse graph to the number of edges in the original graph), is kept between 20\% to 40\% for each stage in order to make the results comparable. For the purpose of our study, we maintain a level span of 3.

\paragraph{Datasets} 
We experiment with 18 real-world networks (see Table \ref{ta2}), which for the purpose of our study we grouped into two groups of social networks, one group of citation networks (CIT) and one group of biological networks (BIO). We split the social networks into 2 groups (SN1, and SN2) - one consisting of Facebook networks, Livejournal and Google+ (general purpose social networks), and the other consisting other consisting of Flickr, Buzznet, Foursquare, Catster, Blogcatalog and Livemocha. The graphs were retrieved from the NetworkRepository \cite{nr-aaai15}, the Koblenz \cite{koblenzCollection}, and the SNAP \cite{snapnets} collections. The size of the networks range between 1 million to 34 million edges.

\begin{table}[H]
	\footnotesize
	\centering
	\caption{Benchmark graphs}
	\label{ta2}
	\begin{tabular}{llccccc}                  
		\hline
		Group & Graph &	$|V|$	&	$|E|$	&	min deg.	& 	max deg. &	avg degree	\\\hline
		Social & fb-indiana	&	29.7K	&	1.3M	&	1	&	1.4K & 87 \\
		Networks 1& fb-texas84	&	36.4K	&	1,6M	&	1	&	6.3K & 87 \\
		(SN1) & fb-uf	&	35.1K	&	1.5M	&	1	&  8.2K & 83 	\\
		& fb-penn94	&	41.5K	&	14M	&	1	&	4.4K & 65	\\
		& livejournal	&	4M	&	27.9M	&	1	&	2.7K & 13 \\
		& google-plus	&	107.6K	&	12.2M	&	1	&	20.1K & 227.4 \\\hline
		Biological & human-gene1	&	22K	&	12M	&	1	&	7.9K	& 1.1K \\
		Networks & human-gene2	&	14K	&	9M	&	1	&	7.2K & 1.3K \\
		(BIO)& Mouse	&	43K	&	14.5M	&	1	&	8K & 670 	\\\hline
		Social  & flickr	&	105K	&	2.3M	&	7	&	5.4K & 43.7 	\\
Networks 2	& buzznet	&	101.2K	&	2.8M	&	1	&	64.3K & 54 \\
		(SN2) & foursquare	&	639K	&	3.2M	&	1	&	106.2 & 10 	\\
		& catster	&	149.7K	&	5.4M	&	1	&	80.6K & 72 	\\
		& blogcatalog	&	88.8K	&	2.1M	&	1	&	9.4K & 47 \\
		& livemocha	&	104.1K	&	2.2M	&	1	&	3K & 42 	\\\hline
		Citation  & ca-cit-Hepth	&	22.9K	&	2.6M	&	1	&	11.9K & 233.38 	\\
		Networks & cit-patent	&	3.7M	&	16.5M	&	1	&	793 & 8.75 \\
		(CIT) & codblp	&	540.5K	&	15.2M	&	1	&	3.3K & 56	\\
		\hline
	\end{tabular}
\end{table}

\paragraph{Methods of Comparison} We studied  various  levels of sparsification while comparing the following  properties of the sparse graph $G_s$, to those in the original graph $G_o$. The single value properties are: (a) \textbf{Diameter} - We measure the ratio of the diameter in $G_o$ to the new diameter in $G_s$ (in plots ``orig diameter/diameter''); (b) \textbf{Number of connected components} - we measure the ratio of the number of connected components in $G_s$ to that in $G_o$ (in plots ``comp/orig comp''); (c) \textbf{Modularity} -  we measure the ratio of modularity in $G_s$ to that of $G_o$ (in plots ``mod/orig mod'', Networkit \cite{networkitRf} provides an implementation of the Louvain method). 
Certain network properties are represented better by their distributions over the nodes. In order to accurately compare the distributions, we use the Spearman rank correlation coefficient. This effectively, reveals how different the sparse graph is from the original in the context of these properties where a correlation value of 1 means they are perfectly correlated and correlation value of 0 means no correlation. The following distributions are compared using the Spearman rank: (a) Node \textbf{betweeenness} centrality; (b)  \textbf{PageRank} centrality; (c) \textbf{Degree distribution}; and (d) \textbf{Clustering coefficient distribution} ($c_i$).  The method changes slightly in comparing node betweenness centrality. Considering that the cost of computing betweenness for large graphs is very expensive, we make use of an approximate method provided by Networkit. However, to ensure accuracy we compute this 10 times and take average of the positional rankings and then compute the Spearman rank correlation.

\subsection{Single-level Algorithm}\label{single-level-results}
\begin{figure}[H]
\hspace{1cm}
\begin{tabular}{m{0cm} m{1cm} m{3.6cm} m{3.6cm} m{3.6cm}}
	\centering
\multirow{8}{*}{\hspace{-2cm}\includegraphics[angle=90,page=25]{results/plots/regular/fb-indiana_fb-uf_fb-texas84_fb-penn94_gplus_lj.pdf}} & &
\centering $\delta$-weak & \centering Mixture & \centering $\delta$-strong \arraybackslash \\
& \raisebox{.25\height}{\rotatebox{90}{\parbox{1.9cm}{Betweenness\\Centrality}}} &
\includegraphics[width=\linewidth, page=1]{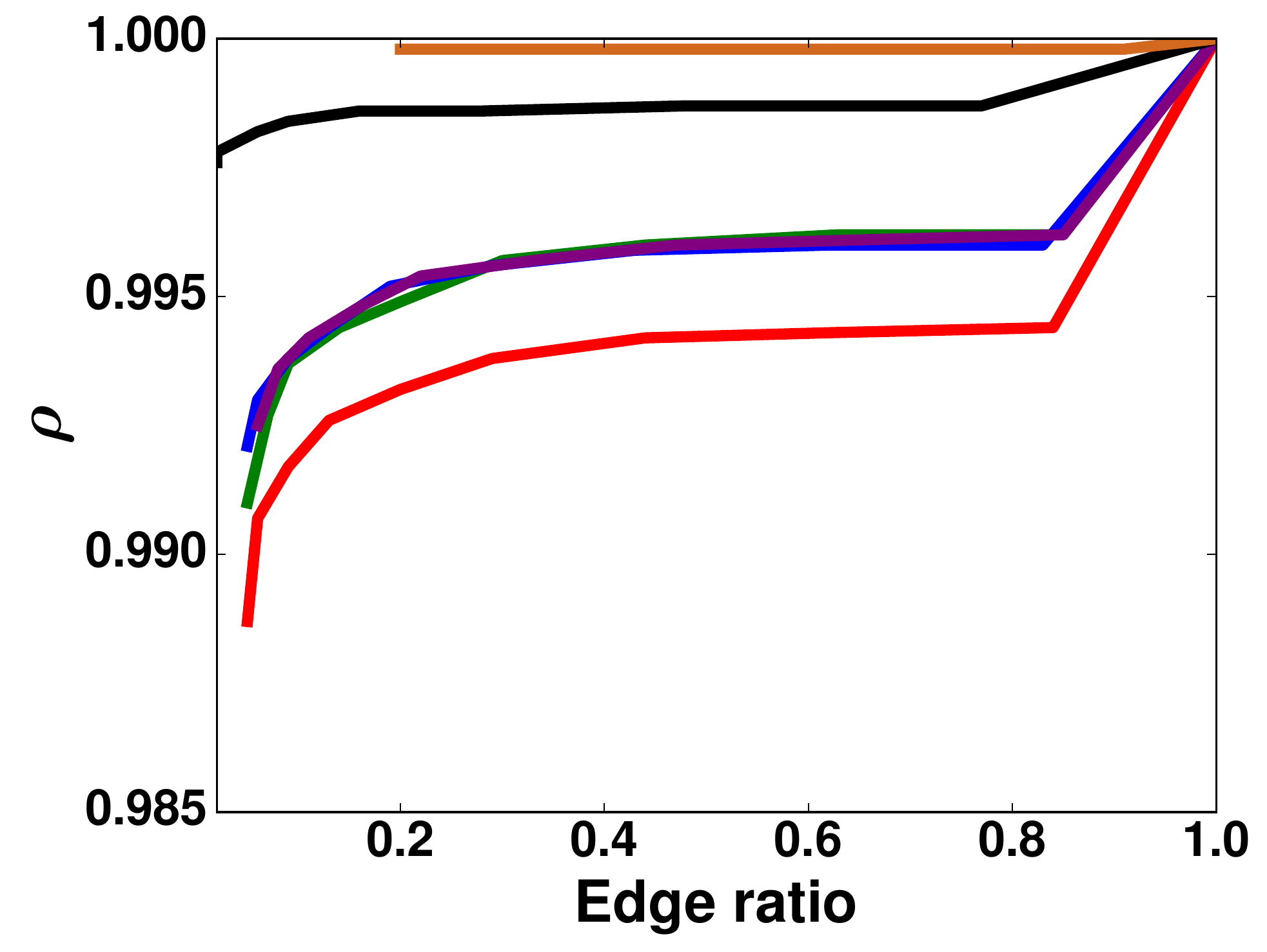} &
\includegraphics[width=\linewidth, page=2]{results/plots/regular/fb-indiana_fb-uf_fb-texas84_fb-penn94_gplus_lj.pdf} &
\includegraphics[width=\linewidth, page=3]{results/plots/regular/fb-indiana_fb-uf_fb-texas84_fb-penn94_gplus_lj.pdf}\\
& \raisebox{.25\height}{\rotatebox{90}{\parbox{1.9cm}{Clustering\\Coefficient\\Distribution}}} &
\includegraphics[width=\linewidth, page=7]{results/plots/regular/fb-indiana_fb-uf_fb-texas84_fb-penn94_gplus_lj.pdf} &
\includegraphics[width=\linewidth, page=8]{results/plots/regular/fb-indiana_fb-uf_fb-texas84_fb-penn94_gplus_lj.pdf} &
\includegraphics[width=\linewidth, page=9]{results/plots/regular/fb-indiana_fb-uf_fb-texas84_fb-penn94_gplus_lj.pdf}\\
& \raisebox{.25\height}{\rotatebox{90}{\parbox{1.9cm}{Degree\\Centrality}}} &
\includegraphics[width=\linewidth, page=10]{results/plots/regular/fb-indiana_fb-uf_fb-texas84_fb-penn94_gplus_lj.pdf} &
\includegraphics[width=\linewidth, page=11]{results/plots/regular/fb-indiana_fb-uf_fb-texas84_fb-penn94_gplus_lj.pdf} &
\includegraphics[width=\linewidth, page=12]{results/plots/regular/fb-indiana_fb-uf_fb-texas84_fb-penn94_gplus_lj.pdf}\\
& \raisebox{.25\height}{\rotatebox{90}{\parbox{1.9cm}{Diameter}}} &
\includegraphics[width=\linewidth, page=13]{results/plots/regular/fb-indiana_fb-uf_fb-texas84_fb-penn94_gplus_lj.pdf} &
\includegraphics[width=\linewidth, page=14]{results/plots/regular/fb-indiana_fb-uf_fb-texas84_fb-penn94_gplus_lj.pdf} &
\includegraphics[width=\linewidth, page=15]{results/plots/regular/fb-indiana_fb-uf_fb-texas84_fb-penn94_gplus_lj.pdf}\\
& \raisebox{.25\height}{\rotatebox{90}{\parbox{1.9cm}{Components}}} &
\includegraphics[width=\linewidth, page=16]{results/plots/regular/fb-indiana_fb-uf_fb-texas84_fb-penn94_gplus_lj.pdf} &
\includegraphics[width=\linewidth, page=17]{results/plots/regular/fb-indiana_fb-uf_fb-texas84_fb-penn94_gplus_lj.pdf} &
\includegraphics[width=\linewidth, page=18]{results/plots/regular/fb-indiana_fb-uf_fb-texas84_fb-penn94_gplus_lj.pdf}\\
& \raisebox{.25\height}{\rotatebox{90}{\parbox{1.9cm}{Modularity}}} &
\includegraphics[width=\linewidth, page=19]{results/plots/regular/fb-indiana_fb-uf_fb-texas84_fb-penn94_gplus_lj.pdf} &
\includegraphics[width=\linewidth, page=20]{results/plots/regular/fb-indiana_fb-uf_fb-texas84_fb-penn94_gplus_lj.pdf} &
\includegraphics[width=\linewidth, page=21]{results/plots/regular/fb-indiana_fb-uf_fb-texas84_fb-penn94_gplus_lj.pdf}\\
& \raisebox{.25\height}{\rotatebox{90}{\parbox{1.9cm}{Pagerank}}} &
\includegraphics[width=\linewidth, page=22]{results/plots/regular/fb-indiana_fb-uf_fb-texas84_fb-penn94_gplus_lj.pdf} &
\includegraphics[width=\linewidth, page=23]{results/plots/regular/fb-indiana_fb-uf_fb-texas84_fb-penn94_gplus_lj.pdf} &
\includegraphics[width=\linewidth, page=24]{results/plots/regular/fb-indiana_fb-uf_fb-texas84_fb-penn94_gplus_lj.pdf}\\
\end{tabular}
\caption{Social Networks 1}\label{fig:soc1}
\end{figure}

\begin{figure}[H]
\footnotesize
\hspace{1cm}
\begin{tabular}{m{0cm} m{1cm} m{3.6cm} m{3.6cm} m{3.6cm}}
	\centering
\multirow{8}{*}{\hspace{-2cm}\includegraphics[angle=90,page=25]{results/plots/regular/flickr_foursquare_catster_buzznet_blogcatalog_livemocha.pdf}} & &
\centering $\delta$-weak & \centering Mixture & \centering $\delta$-strong \arraybackslash \\
& \raisebox{.25\height}{\rotatebox{90}{\parbox{1.9cm}{Betweenness\\Centrality}}} &
\includegraphics[width=\linewidth, page=1]{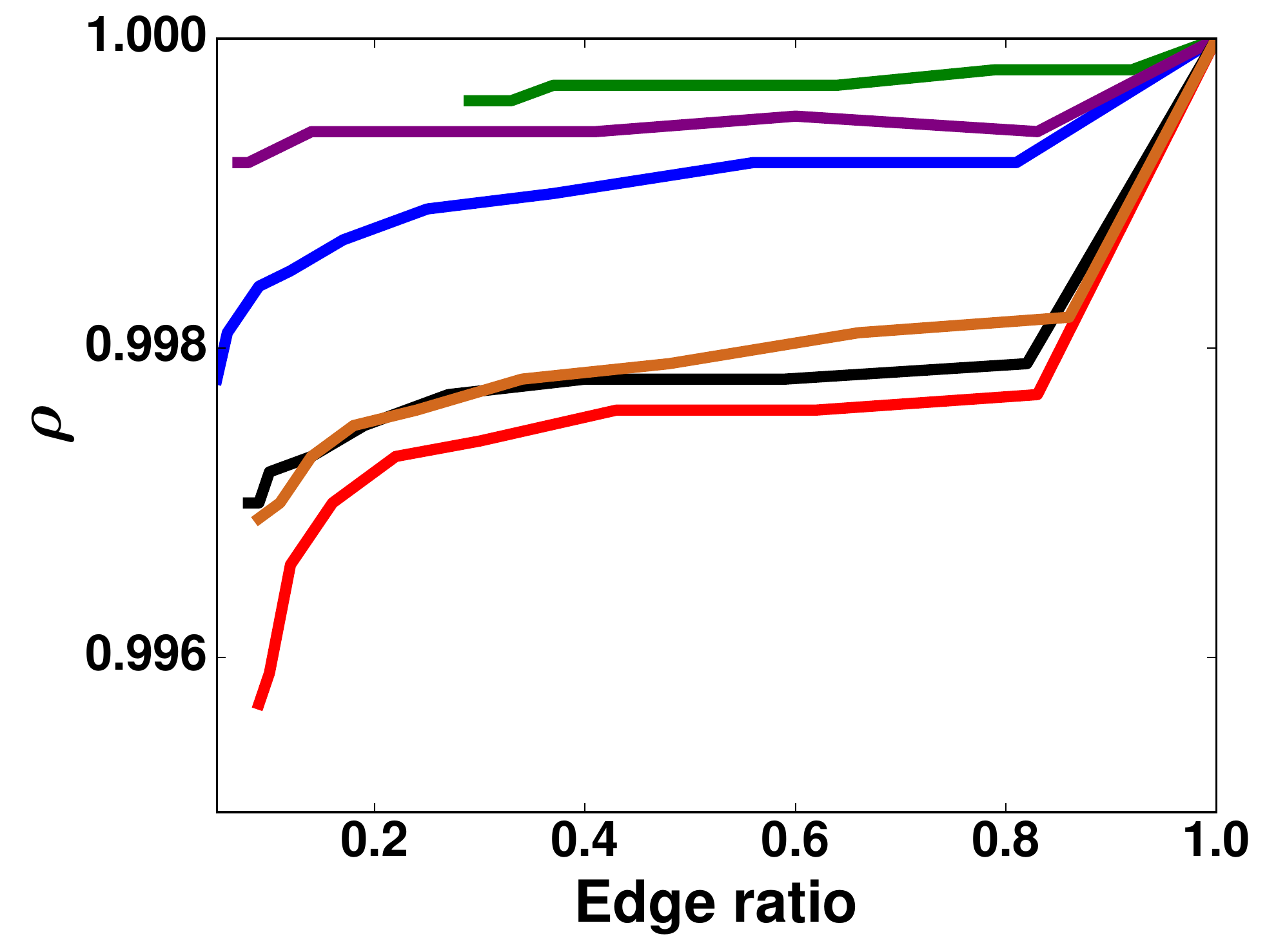} &
\includegraphics[width=\linewidth, page=2]{results/plots/regular/flickr_foursquare_catster_buzznet_blogcatalog_livemocha.pdf} &
\includegraphics[width=\linewidth, page=3]{results/plots/regular/flickr_foursquare_catster_buzznet_blogcatalog_livemocha.pdf}\\
& \raisebox{.25\height}{\rotatebox{90}{\parbox{1.9cm}{Clustering\\Coefficient\\Distribution}}} &
\includegraphics[width=\linewidth, page=7]{results/plots/regular/flickr_foursquare_catster_buzznet_blogcatalog_livemocha.pdf} &
\includegraphics[width=\linewidth, page=8]{results/plots/regular/flickr_foursquare_catster_buzznet_blogcatalog_livemocha.pdf} &
\includegraphics[width=\linewidth, page=9]{results/plots/regular/flickr_foursquare_catster_buzznet_blogcatalog_livemocha.pdf}\\
& \raisebox{.25\height}{\rotatebox{90}{\parbox{1.9cm}{Degree\\Centrality}}} &
\includegraphics[width=\linewidth, page=10]{results/plots/regular/flickr_foursquare_catster_buzznet_blogcatalog_livemocha.pdf} &
\includegraphics[width=\linewidth, page=11]{results/plots/regular/flickr_foursquare_catster_buzznet_blogcatalog_livemocha.pdf} &
\includegraphics[width=\linewidth, page=12]{results/plots/regular/flickr_foursquare_catster_buzznet_blogcatalog_livemocha.pdf}\\
& \raisebox{.25\height}{\rotatebox{90}{\parbox{1.9cm}{Diameter}}} &
\includegraphics[width=\linewidth, page=13]{results/plots/regular/flickr_foursquare_catster_buzznet_blogcatalog_livemocha.pdf} &
\includegraphics[width=\linewidth, page=14]{results/plots/regular/flickr_foursquare_catster_buzznet_blogcatalog_livemocha.pdf} &
\includegraphics[width=\linewidth, page=15]{results/plots/regular/flickr_foursquare_catster_buzznet_blogcatalog_livemocha.pdf}\\
& \raisebox{.25\height}{\rotatebox{90}{\parbox{1.9cm}{Components}}} &
\includegraphics[width=\linewidth, page=16]{results/plots/regular/flickr_foursquare_catster_buzznet_blogcatalog_livemocha.pdf} &
\includegraphics[width=\linewidth, page=17]{results/plots/regular/flickr_foursquare_catster_buzznet_blogcatalog_livemocha.pdf} &
\includegraphics[width=\linewidth, page=18]{results/plots/regular/flickr_foursquare_catster_buzznet_blogcatalog_livemocha.pdf}\\
& \raisebox{.25\height}{\rotatebox{90}{\parbox{1.9cm}{Modularity}}} &
\includegraphics[width=\linewidth, page=19]{results/plots/regular/flickr_foursquare_catster_buzznet_blogcatalog_livemocha.pdf} &
\includegraphics[width=\linewidth, page=20]{results/plots/regular/flickr_foursquare_catster_buzznet_blogcatalog_livemocha.pdf} &
\includegraphics[width=\linewidth, page=21]{results/plots/regular/flickr_foursquare_catster_buzznet_blogcatalog_livemocha.pdf}\\
& \raisebox{.25\height}{\rotatebox{90}{\parbox{1.9cm}{Pagerank}}} &
\includegraphics[width=\linewidth, page=22]{results/plots/regular/flickr_foursquare_catster_buzznet_blogcatalog_livemocha.pdf} &
\includegraphics[width=\linewidth, page=23]{results/plots/regular/flickr_foursquare_catster_buzznet_blogcatalog_livemocha.pdf} &
\includegraphics[width=\linewidth, page=24]{results/plots/regular/flickr_foursquare_catster_buzznet_blogcatalog_livemocha.pdf}\\
\end{tabular}
\caption{Social Networks 2}\label{fig:soc2}
\end{figure}

\begin{figure}[H]
\footnotesize
\hspace{1cm}
\begin{tabular}{m{0cm} m{1cm} m{3.6cm} m{3.6cm} m{3.6cm}}
	\centering
\multirow{8}{*}{\hspace{-2cm}\includegraphics[angle=90,page=25]{results/plots/regular/ca-cit-Hepth_codblp_cit-patent.pdf}} & &
\centering $\delta$-weak & \centering Mixture & \centering $\delta$-strong \arraybackslash \\
& \raisebox{.25\height}{\rotatebox{90}{\parbox{1.9cm}{Betweenness\\Centrality}}} &
\includegraphics[width=\linewidth, page=1]{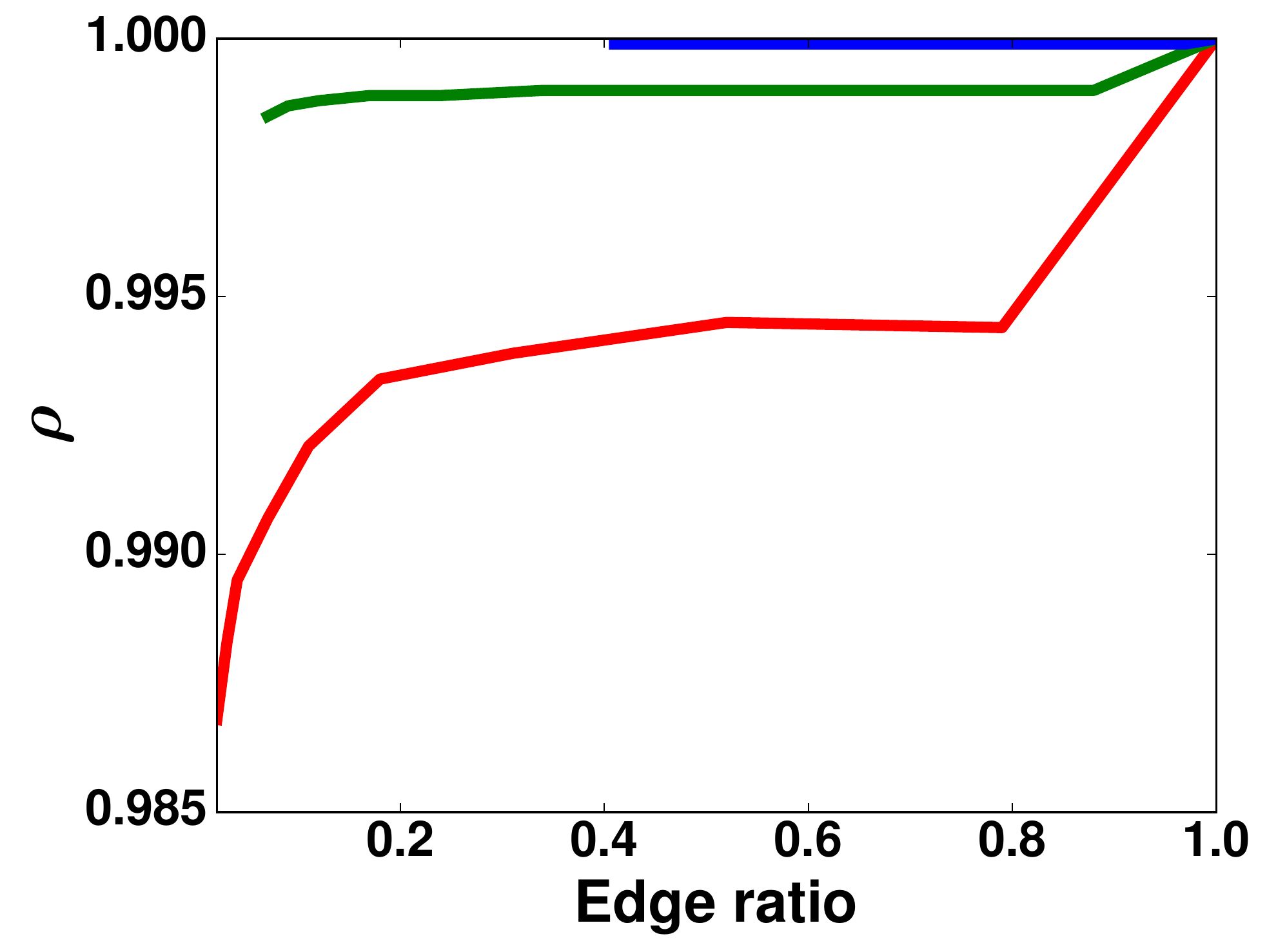} &
\includegraphics[width=\linewidth, page=2]{results/plots/regular/ca-cit-Hepth_codblp_cit-patent.pdf} &
\includegraphics[width=\linewidth, page=3]{results/plots/regular/ca-cit-Hepth_codblp_cit-patent.pdf}\\
& \raisebox{.25\height}{\rotatebox{90}{\parbox{1.9cm}{Clustering\\Coefficient\\Distribution}}} &
\includegraphics[width=\linewidth, page=7]{results/plots/regular/ca-cit-Hepth_codblp_cit-patent.pdf} &
\includegraphics[width=\linewidth, page=8]{results/plots/regular/ca-cit-Hepth_codblp_cit-patent.pdf} &
\includegraphics[width=\linewidth, page=9]{results/plots/regular/ca-cit-Hepth_codblp_cit-patent.pdf}\\
& \raisebox{.25\height}{\rotatebox{90}{\parbox{1.9cm}{Degree\\Centrality}}} &
\includegraphics[width=\linewidth, page=10]{results/plots/regular/ca-cit-Hepth_codblp_cit-patent.pdf} &
\includegraphics[width=\linewidth, page=11]{results/plots/regular/ca-cit-Hepth_codblp_cit-patent.pdf} &
\includegraphics[width=\linewidth, page=12]{results/plots/regular/ca-cit-Hepth_codblp_cit-patent.pdf}\\
& \raisebox{.25\height}{\rotatebox{90}{\parbox{1.9cm}{Diameter}}} &
\includegraphics[width=\linewidth, page=13]{results/plots/regular/ca-cit-Hepth_codblp_cit-patent.pdf} &
\includegraphics[width=\linewidth, page=14]{results/plots/regular/ca-cit-Hepth_codblp_cit-patent.pdf} &
\includegraphics[width=\linewidth, page=15]{results/plots/regular/ca-cit-Hepth_codblp_cit-patent.pdf}\\
& \raisebox{.25\height}{\rotatebox{90}{\parbox{1.9cm}{Components}}} &
\includegraphics[width=\linewidth, page=16]{results/plots/regular/ca-cit-Hepth_codblp_cit-patent.pdf} &
\includegraphics[width=\linewidth, page=17]{results/plots/regular/ca-cit-Hepth_codblp_cit-patent.pdf} &
\includegraphics[width=\linewidth, page=18]{results/plots/regular/ca-cit-Hepth_codblp_cit-patent.pdf}\\
& \raisebox{.25\height}{\rotatebox{90}{\parbox{1.9cm}{Modularity}}} &
\includegraphics[width=\linewidth, page=19]{results/plots/regular/ca-cit-Hepth_codblp_cit-patent.pdf} &
\includegraphics[width=\linewidth, page=20]{results/plots/regular/ca-cit-Hepth_codblp_cit-patent.pdf} &
\includegraphics[width=\linewidth, page=21]{results/plots/regular/ca-cit-Hepth_codblp_cit-patent.pdf}\\
& \raisebox{.25\height}{\rotatebox{90}{\parbox{1.9cm}{Pagerank}}} &
\includegraphics[width=\linewidth, page=22]{results/plots/regular/ca-cit-Hepth_codblp_cit-patent.pdf} &
\includegraphics[width=\linewidth, page=23]{results/plots/regular/ca-cit-Hepth_codblp_cit-patent.pdf} &
\includegraphics[width=\linewidth, page=24]{results/plots/regular/ca-cit-Hepth_codblp_cit-patent.pdf}\\
\end{tabular}
\caption{Citation Networks}\label{fig:cit}
\end{figure}

\begin{figure}[H]
\footnotesize
\hspace{1cm}
\begin{tabular}{m{0cm} m{1cm} m{3.6cm} m{3.6cm} m{3.6cm}}
	\centering
\multirow{8}{*}{\hspace{-2cm}\includegraphics[angle=90,page=25]{results/plots/regular/human1_human2_mouse.pdf}} & &
\centering $\delta$-weak & \centering Mixture & \centering $\delta$-strong \arraybackslash \\
& \raisebox{.25\height}{\rotatebox{90}{\parbox{1.9cm}{Betweenness\\Centrality}}} &
\includegraphics[width=\linewidth, page=1]{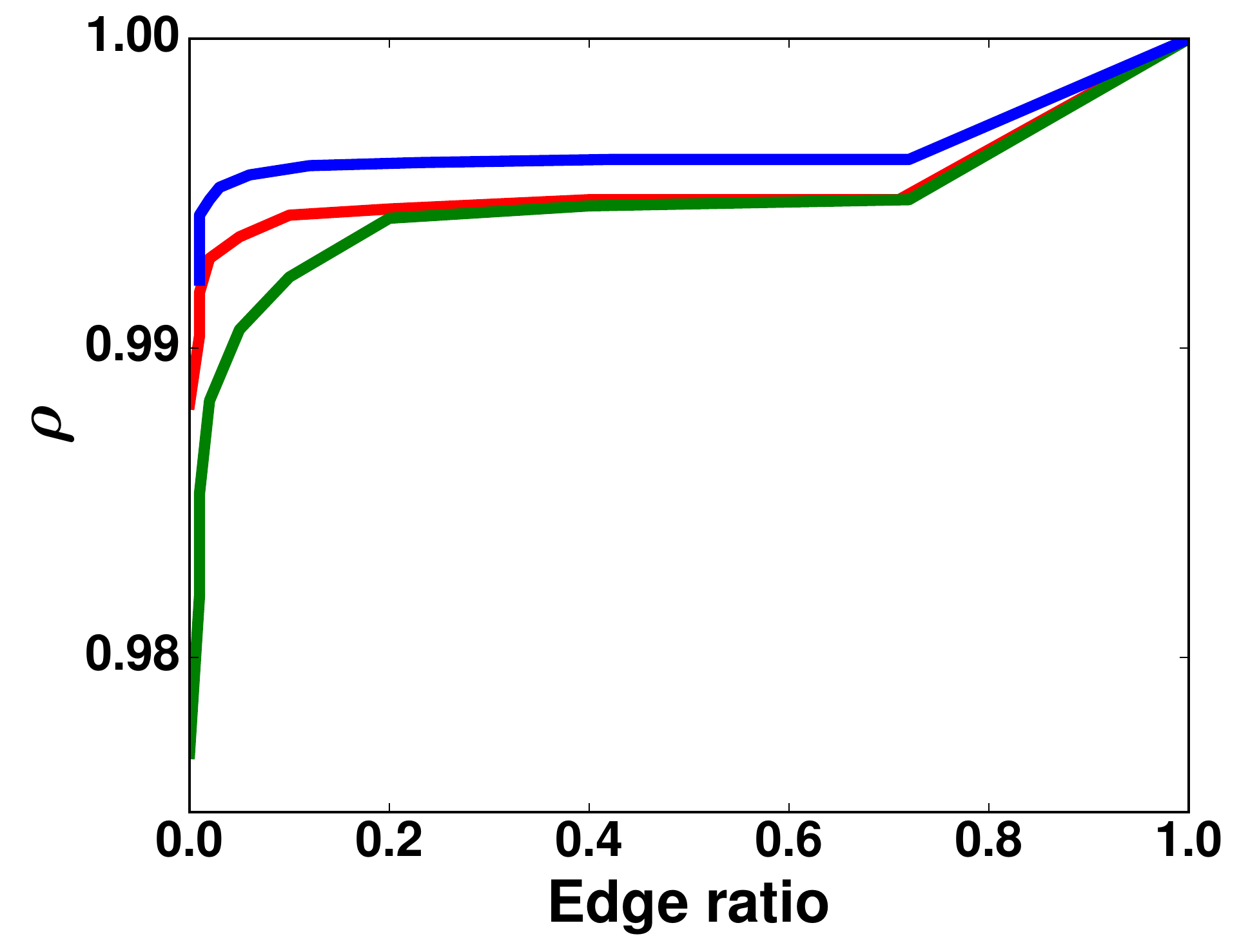} &
\includegraphics[width=\linewidth, page=2]{results/plots/regular/human1_human2_mouse.pdf} &
\includegraphics[width=\linewidth, page=3]{results/plots/regular/human1_human2_mouse.pdf}\\
& \raisebox{.25\height}{\rotatebox{90}{\parbox{1.9cm}{Clustering\\Coefficient\\Distribution}}} &
\includegraphics[width=\linewidth, page=7]{results/plots/regular/human1_human2_mouse.pdf} &
\includegraphics[width=\linewidth, page=8]{results/plots/regular/human1_human2_mouse.pdf} &
\includegraphics[width=\linewidth, page=9]{results/plots/regular/human1_human2_mouse.pdf}\\
& \raisebox{.25\height}{\rotatebox{90}{\parbox{1.9cm}{Degree\\Centrality}}} &
\includegraphics[width=\linewidth, page=10]{results/plots/regular/human1_human2_mouse.pdf} &
\includegraphics[width=\linewidth, page=11]{results/plots/regular/human1_human2_mouse.pdf} &
\includegraphics[width=\linewidth, page=12]{results/plots/regular/human1_human2_mouse.pdf}\\
& \raisebox{.25\height}{\rotatebox{90}{\parbox{1.9cm}{Diameter}}} &
\includegraphics[width=\linewidth, page=13]{results/plots/regular/human1_human2_mouse.pdf} &
\includegraphics[width=\linewidth, page=14]{results/plots/regular/human1_human2_mouse.pdf} &
\includegraphics[width=\linewidth, page=15]{results/plots/regular/human1_human2_mouse.pdf}\\
& \raisebox{.25\height}{\rotatebox{90}{\parbox{1.9cm}{Components}}} &
\includegraphics[width=\linewidth, page=16]{results/plots/regular/human1_human2_mouse.pdf} &
\includegraphics[width=\linewidth, page=17]{results/plots/regular/human1_human2_mouse.pdf} &
\includegraphics[width=\linewidth, page=18]{results/plots/regular/human1_human2_mouse.pdf}\\
& \raisebox{.25\height}{\rotatebox{90}{\parbox{1.9cm}{Modularity}}} &
\includegraphics[width=\linewidth, page=19]{results/plots/regular/human1_human2_mouse.pdf} &
\includegraphics[width=\linewidth, page=20]{results/plots/regular/human1_human2_mouse.pdf} &
\includegraphics[width=\linewidth, page=21]{results/plots/regular/human1_human2_mouse.pdf}\\
& \raisebox{.25\height}{\rotatebox{90}{\parbox{1.9cm}{Pagerank}}} &
\includegraphics[width=\linewidth, page=22]{results/plots/regular/human1_human2_mouse.pdf} &
\includegraphics[width=\linewidth, page=23]{results/plots/regular/human1_human2_mouse.pdf} &
\includegraphics[width=\linewidth, page=24]{results/plots/regular/human1_human2_mouse.pdf}\\
\end{tabular}
\caption{Biological Networks}\label{fig:bio}
\end{figure}

The single-level algorithm was tested with both unnormalized and normalized algebraic distances. The  results for unnormalized algebraic distance are presented in Figures \ref{fig:soc1}, \ref{fig:soc2}, \ref{fig:cit}, and \ref{fig:bio} for groups SN1, SN2, CIT, and BIO, respectively. (The  results for the normalized algebraic distance can be found in Appendix \ref{App:AppendixA}.) In each figure, 3 columns, and 7 rows of plots are presented. In all 4 figures: (a) each column corresponds to the type of filtering, i.e., to the types of edges that retain after sparsification; (b) each row corresponds to the type of comparison. Each plot contains several colored curves that correspond to the respective graphs (see vertical legend). One point in each curve corresponds to an average of the measured comparison method over 10 runs  for the corresponding edge ratio in each. The x- and y-axes correspond to the sparsification ratio and method of comparison, respectively. In the y-axis of betweenness, PageRank, degree, and clustering coefficient distribution centralities, the Spearman rank is denoted by $\rho$. For example, we examine the behavior of the degrees in social network Google+ in SN1 when $\delta$-strong edges retain after sparsification. In Figure \ref{fig:soc1}, we find a row ``Degree centrality'' (row 3). The results for retaining $\delta$-strong edges are found in the third column. The black curve corresponds to Google+, where each point is an average of 10 runs.\\
\noindent \emph{Note: Most curves do not reach a visible zero of the x-axis. This is because the sparsification is interrupted when the number of edges becomes less than the number of nodes.}



\paragraph{$\delta$-weak edges} Plots labelled as $\delta$-weak (column 1) are results obtained by retaining only weak edges,  when $\delta$-weak edges are preferred during sparsification (i.e, $\delta$-strong edges are deleted). In this type of sparsification, we expect that sparsification of the local structure will mostly dominate the sparsification of the global structure. Indeed, we observe that properties (such as the betweenness centrality, diameter, and the number of components) that heavily depend on usually limited number of long-range weak connections are well preserved. 



\paragraph{$\delta$-strong} Plots labelled as $\delta$-strong (column 3) are results obtained by retaining $\delta$-strong edges and removing $\delta$-weak edges. By preferring $\delta$-strong edges during sparsification, we attempt to preserve properties that depends on the local structure of the graph. Such properties as clustering coefficient, pagerank and degree centrality survive sparsification better when this method is used. In particular, we can observe that the clustering coefficient (which is in many cases the reason for a strong community structure) is preserved at the level of $\approx 75\%$ in SN1 when 70\% of edges are removed (instead of $\approx 40\%$ for $\delta$-weak sparsification). A similar phenomena is observed in BIO. It is interesting to note that in SN2, in comparison to the $\delta$-weak sparsification, the changes in the clustering coefficient are not significant.



\paragraph{Mixture sparsification} In plots labelled as mixed, we maintain a balance between the $\delta$-weak and $\delta$-strong types of sparsification by preferring ensuring that both are sparsified. For such properties as the betweenness centrality, PageRank and degree centrality, the results are better for up to 20\% sparsification ratio when compared to selecting either weak or strong edges. For such properties as the clustering coefficient, modularity, diameter and connected components, retaining both weak and strong edges provides results that is in between that produced by weak or strong edges sparsification.

\newpage
\footnotesize
\normalsize

\footnotesize
\normalsize

\footnotesize
\normalsize

\footnotesize
\normalsize

\begin{figure}[h]
\footnotesize
\hspace{1cm}
\begin{tabular}{m{0.5cm} m{3.8cm} m{3.8cm} m{3.8cm}}
\multirow{4}{*}{\hspace{-0.2cm}\includegraphics[angle=90,page=8]{results/compare_benchmark/plots/compare-ld.pdf}} & & & \arraybackslash \\
&\includegraphics[width=\linewidth, page=1]{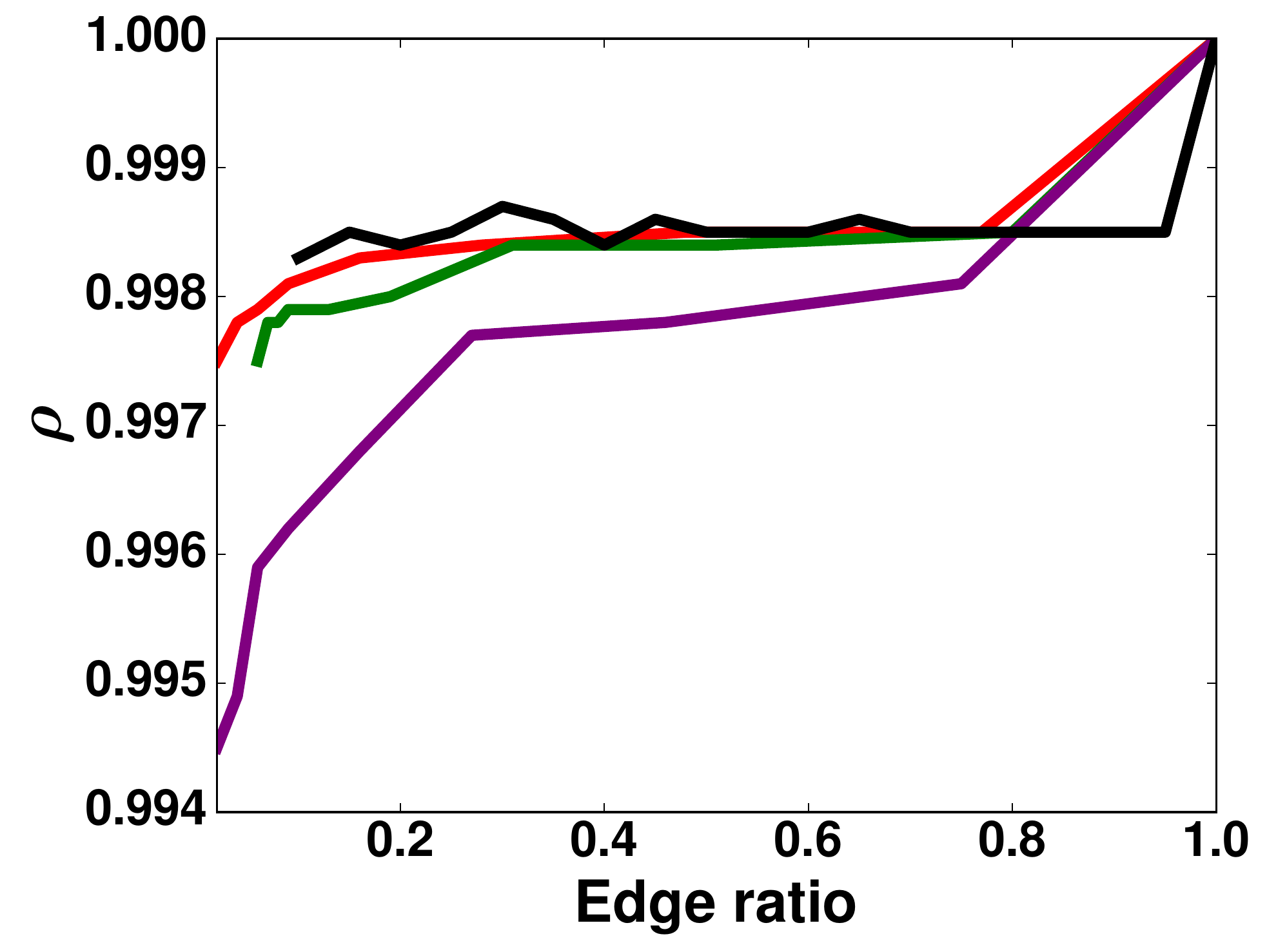} &
\includegraphics[width=\linewidth, page=2]{results/compare_benchmark/plots/compare-ld.pdf} &
\includegraphics[width=\linewidth, page=3]{results/compare_benchmark/plots/compare-ld.pdf}\\
&\centering Betweenness Centrality & \centering Clustering Coefficient \\ Distribution & \centering Degree Centrality \arraybackslash \\
&\includegraphics[width=\linewidth, page=4]{results/compare_benchmark/plots/compare-ld.pdf}&
\includegraphics[width=\linewidth, page=5]{results/compare_benchmark/plots/compare-ld.pdf} &
\includegraphics[width=\linewidth, page=6]{results/compare_benchmark/plots/compare-ld.pdf}\\
&\centering Diameter & \centering Connected Components & \centering Modularity \arraybackslash \\
&\includegraphics[width=\linewidth, page=7]{results/compare_benchmark/plots/compare-ld.pdf} & &\\
&\centering PageRank &  &  \\
\end{tabular}
\caption{Comparison of LD and single-level algebraic distance methods.}\label{ld-sl}
\end{figure}

\paragraph{Comparing with Local Degree} In the introduction, we mentioned the Local Degree method (LD) \cite{lindner2015structure} which favors the retention of edges participating in hubs (nodes with high degree). In order to compare our method to LD, we ran the single level algorithm for retaining weak edges, strong edges and a mixture of both on the Google+ graph (google-plus in Table \ref{ta2}). Same set of network properties discussed earlier in this section were used for comparison. For betweenness centrality, degree centrality, local clustering coefficient and PageRank, we plot the Spearman rank correlation against the edge ratio. Figure \ref{ld-sl} shows the plots of $\delta$-weak, $\delta$-strong, mixed and local degree(LD) for each property. The results are similar for betweenness centrality, degree centrality and PageRank. However, for such properties as modularity and clustering coefficient, the algebraic distance performs better than LD especially when sparsification is aggressive. The $\delta$-weak sparsification preserves the diameter slightly bettern than LD while the $\delta$-strong method did not perform well on it and on the number of components. We note that the LD method was comprehensively studied on the Facebook networks only. Four Facebook networks in SN1 demonstrate similar performance with both methods. The Google+ network has exceptionally high clustering coefficient (0.52 vs. 0.23 in Facebook networks) and smaller diameter (6 vs. 8 in Facebook networks) which are more difficult to preserve if the method does not distinguish between local- and global-range connections.



\footnotesize
\normalsize


\subsection{Multi-Level Results}\label{sub:MultiscaleResults}
The purpose of the multilevel approach is to extend the general sparsification framework to enable highly controllable sparsification of bundles of edges at multiple coarse-grained resolutions. Similar to the  single-level experiments, we group the networks into 4 groups. Tables \ref{ms_sn1}, \ref{ms_sn2}, \ref{ms_bio}, and \ref{ms_cit} show the results of the multilevel algorithm for sets SN1, SN2, BIO and CIT, respectively. 
The 4 major column sections in the aforementioned tables consists of the graph name, the levels' configuration for which sparsifcation is tested, the number of edges at each setting, and the network properties that we study. The properties column consist of the following properties: a) CC - clustering coefficient b) D - Diameter of the graph, c) Q - Modularity of the graph, d) $\Gamma$ - the number of components in the network, e) $BC_{\rho}$ - Spearman rank correlation for betweenness centrality, f) $PR_{\rho}$ - Spearman rank correlation of Pagerank, g) $DC_{\rho}$ - Spearman rank correlation of degree centrality, and h) $CC_{\rho}$ - Spearman rank correlation of the clustering coefficient. Correlation here represents the correlation between the original graph and the sparse graph. 
The "Level" column contains the sparsification settings at different levels, where G0 represents the original graph, G1 is the graph with sparsification only at the coarsest levels, G2 is the graph with sparsification only at the middle levels and G3 represents the graph with sparsification at the fine levels. In order to keep the results comparable, we keep the sparsification ratio between 20\% to 40\%. The sparsification parameter is obtained by a binary search fitting algorithm. 
Note that we do not compare the sparse graphs (G1, G2, G3) to themselves but only with the fine graph G0. The parameter setting of a coarsening was similar to one described in \cite{safro:relaxml} with interpolation order 1.

\begin{table}[h]
	\centering
	\resizebox{0.9\textwidth}{!}{%
		\begin{tabular}{ |l|l|l|l|l|l|l|l|l|l|l| }
		\hline
		\multirow{2}{*}{\textbf{Graph Name}} & \multirow{2}{*}{\textbf{Level}} & \multirow{2}{*}{\textbf{$|E|$}} & \multicolumn{8}{c|}{ \textbf{Properties} } \\
		\cline{4 - 11}
		& & & CC & D & Q & $\Gamma$ & $BC_{\rho}$ & $PR_{\rho}$ & $DC_{\rho}$ & $CC_{\rho}$\\
		\hline\multirow{4}{*}{fb-indiana}& G0 & 1.3M & 0.21 & 8.0 & 0.45 & 1 & 1.0 & 1.0 & 1.0 & 1.0\\
& G1 & 361.4K & 0.31 & 13.0 & 0.93 & 18 & 1.0 & 0.84 & 0.83 & 0.64\\
& G2 & 349.8K & 0.14 & 10.0 & 0.34 & 8 & 0.99 & 0.89 & 0.89 & 0.57\\
& G3 & 402.7K & 0.04 & 12.0 & 0.3 & 37 & 0.99 & 0.93 & 0.95 & 0.03\\
\hline
\multirow{4}{*}{fb-texas84}& G0 & 1.6M & 0.2 & 7.0 & 0.38 & 1 & 1.0 & 1.0 & 1.0 & 1.0\\
& G1 & 374.7K & 0.29 & 16.0 & 0.92 & 16 & 1.0 & 0.86 & 0.82 & 0.57\\
& G2 & 574.8K & 0.13 & 9.0 & 0.31 & 5 & 0.99 & 0.94 & 0.94 & 0.66\\
& G3 & 521.9K & 0.05 & 12.0 & 0.24 & 29 & 1.0 & 0.94 & 0.96 & 0.08\\
\hline
\multirow{4}{*}{fb-uf}& G0 & 1.5M & 0.22 & 8.0 & 0.44 & 1 & 1.0 & 1.0 & 1.0 & 1.0\\
& G1 & 423.4K & 0.24 & 12.0 & 0.61 & 3 & 0.99 & 0.88 & 0.88 & 0.58\\
& G2 & 425.1K & 0.16 & 11.0 & 0.37 & 15 & 1.0 & 0.9 & 0.9 & 0.6\\
& G3 & 418.9K & 0.05 & 11.0 & 0.27 & 57 & 1.0 & 0.92 & 0.95 & -0.01\\
\hline
\multirow{4}{*}{fb-penn94}& G0 & 1.4M & 0.22 & 8.0 & 0.48 & 1 & 1.0 & 1.0 & 1.0 & 1.0\\
& G1 & 351.7K & 0.33 & 16.0 & 0.95 & 16 & 1.0 & 0.85 & 0.8 & 0.57\\
& G2 & 463.4K & 0.16 & 11.0 & 0.38 & 23 & 1.0 & 0.89 & 0.9 & 0.6\\
& G3 & 365.6K & 0.04 & 14.0 & 0.3 & 122 & 1.0 & 0.89 & 0.92 & -0.09\\
\hline
\multirow{4}{*}{livejournal}& G0 & 34.7M & 0.35 & 21.0 & 0.75 & 1 & 1.0 & 1.0 & 1.0 & 1.0\\
& G1 & 13.6M & 0.47 & 72.0 & 1.0 & 1.7K & 1.0 & 0.85 & 0.81 & 0.75\\
& G2 & 13.4M & 0.39 & 42.0 & 0.8 & 7.1K & 1.0 & 0.91 & 0.87 & 0.76\\
& G3 & 8.2M & 0.02 & 30.0 & 0.72 & 316.9K & 1.0 & 0.67 & 0.73 & 0.37\\
\hline
\multirow{4}{*}{gplus}& G0 & 12.2M & 0.52 & 6.0 & 0.47 & 1 & 1.0 & 1.0 & 1.0 & 1.0\\
& G1 & 3.6M & 0.54 & 14.0 & 0.89 & 15 & 1.0 & 0.91 & 0.9 & 0.59\\
& G2 & 3.0M & 0.42 & 12.0 & 0.38 & 18 & 1.0 & 0.9 & 0.88 & 0.57\\
& G3 & 3.3M & 0.15 & 19.0 & 0.26 & 905 & 1.0 & 0.79 & 0.88 & -0.26\\
\hline
\end{tabular}

	}
	\caption{Multiscale results for social networks 1 (SN1) graphs}\label{ms_sn1}
\end{table}

\begin{table}[H]
	\centering
	\footnotesize
	\resizebox{0.9\textwidth}{!}{%
		\begin{tabular}{ |l|l|l|l|l|l|l|l|l|l|l| }
		\hline
		\multirow{2}{*}{\textbf{Graph Name}} & \multirow{2}{*}{\textbf{Level}} & \multirow{2}{*}{\textbf{$|E|$}} & \multicolumn{8}{c|}{ \textbf{Properties} } \\
		\cline{4 - 11}
		& & & CC & D & Q & $\Gamma$ & $BC_{\rho}$ & $PR_{\rho}$ & $DC_{\rho}$ & $CC_{\rho}$\\
		\hline\multirow{4}{*}{flickr}& G0 & 2.3M & 0.09 & 9.0 & 0.67 & 83 & 1.0 & 1.0 & 1.0 & 1.0\\
& G1 & 921.0K & 0.15 & 35.0 & 0.91 & 144 & 1.0 & 0.6 & 0.62 & 0.78\\
& G2 & 496.1K & 0.12 & 18.0 & 0.84 & 134 & 1.0 & 0.71 & 0.73 & 0.8\\
& G3 & 634.9K & 0.04 & 25.0 & 0.55 & 5.8K & 1.0 & 0.64 & 0.77 & 0.74\\
\hline
\multirow{4}{*}{buzznet}& G0 & 2.8M & 0.25 & 5.0 & 0.31 & 1 & 1.0 & 1.0 & 1.0 & 1.0\\
& G1 & 713.8K & 0.26 & 11.0 & 0.5 & 1 & 1.0 & 0.82 & 0.91 & 0.6\\
& G2 & 919.3K & 0.21 & 18.0 & 0.3 & 12 & 1.0 & 0.83 & 0.94 & 0.6\\
& G3 & 666.6K & 0.1 & 16.0 & 0.29 & 239 & 1.0 & 0.75 & 0.89 & 0.28\\
\hline
\multirow{4}{*}{foursquare}& G0 & 3.2M & 0.22 & 4.0 & 0.41 & 1 & 1.0 & 1.0 & 1.0 & 1.0\\
& G1 & 1.1M & 0.17 & 36.0 & 0.94 & 18 & 1.0 & 0.74 & 0.88 & 0.87\\
& G2 & 765.9K & 0.19 & 47.0 & 0.94 & 366 & 1.0 & 0.49 & 0.78 & 0.79\\
& G3 & 1.1M & 0.04 & 14.0 & 0.57 & 10.0K & 1.0 & 0.36 & 0.74 & 0.73\\
\hline
\multirow{4}{*}{catster}& G0 & 5.4M & 0.41 & 10.0 & 0.39 & 281 & 1.0 & 1.0 & 1.0 & 1.0\\
& G1 & 1.3M & 0.31 & 15.0 & 0.69 & 293 & 1.0 & 0.59 & 0.59 & 0.39\\
& G2 & 1.7M & 0.26 & 11.0 & 0.37 & 360 & 1.0 & 0.86 & 0.87 & 0.63\\
& G3 & 1.5M & 0.27 & 14.0 & 0.29 & 1.1K & 1.0 & 0.76 & 0.84 & 0.4\\
\hline
\multirow{4}{*}{blog-catalog}& G0 & 2.1M & 0.46 & 9.0 & 0.32 & 1 & 1.0 & 1.0 & 1.0 & 1.0\\
& G1 & 423.9K & 0.41 & 14.0 & 0.67 & 1 & 1.0 & 0.81 & 0.87 & 0.47\\
& G2 & 570.8K & 0.26 & 11.0 & 0.27 & 9 & 1.0 & 0.85 & 0.89 & 0.44\\
& G3 & 566.1K & 0.18 & 12.0 & 0.23 & 391 & 1.0 & 0.7 & 0.83 & 0.29\\
\hline
\multirow{4}{*}{livemocha}& G0 & 2.2M & 0.06 & 6.0 & 0.36 & 1 & 1.0 & 1.0 & 1.0 & 1.0\\
& G1 & 636.9K & 0.04 & 12.0 & 0.45 & 3 & 1.0 & 0.89 & 0.91 & 0.48\\
& G2 & 869.1K & 0.04 & 10.0 & 0.29 & 8 & 1.0 & 0.9 & 0.91 & 0.49\\
& G3 & 556.7K & 0.02 & 11.0 & 0.29 & 1.4K & 1.0 & 0.77 & 0.88 & 0.38\\
\hline
\end{tabular}

	}
	\caption{Multiscale results for social networks 2(SN2) graphs}\label{ms_sn2}
\end{table}

\begin{table}[H]
	\centering
	\footnotesize
	\resizebox{\textwidth}{!}{%
		\begin{tabular}{ |l|l|l|l|l|l|l|l|l|l|l| }
		\hline
		\multirow{2}{*}{\textbf{Graph Name}} & \multirow{2}{*}{\textbf{Level}} & \multirow{2}{*}{\textbf{$|E|$}} & \multicolumn{8}{c|}{ \textbf{Properties} } \\
		\cline{4 - 11}
		& & & CC & D & Q & $\Gamma$ & $BC_{\rho}$ & $PR_{\rho}$ & $DC_{\rho}$ & $CC_{\rho}$\\
		\hline\multirow{4}{*}{bio-human-gene1}& G0 & 12.3M & 0.63 & 8.0 & 0.38 & 17 & 1.0 & 1.0 & 1.0 & 1.0\\
& G1 & 2.8M & 0.66 & 18.0 & 0.8 & 19 & 0.99 & 0.9 & 0.95 & 0.74\\
& G2 & 4.0M & 0.33 & 9.0 & 0.39 & 22 & 0.99 & 0.91 & 0.91 & 0.71\\
& G3 & 3.9M & 0.06 & 11.0 & 0.33 & 78 & 0.99 & 0.89 & 0.93 & 0.21\\
\hline
\multirow{4}{*}{bio-human-gene2}& G0 & 9.0M & 0.66 & 7.0 & 0.31 & 2 & 1.0 & 1.0 & 1.0 & 1.0\\
& G1 & 2.4M & 0.67 & 7.0 & 0.74 & 15 & 1.0 & 0.87 & 0.86 & 0.74\\
& G2 & 3.1M & 0.64 & 26.0 & 0.52 & 14 & 0.98 & 0.87 & 0.88 & 0.74\\
& G3 & 2.5M & 0.62 & 39.0 & 0.42 & 66 & 0.93 & 0.88 & 0.87 & 0.64\\
\hline
\multirow{4}{*}{bio-mouse-gene}& G0 & 14.5M & 0.53 & 12.0 & 0.62 & 97 & 1.0 & 1.0 & 1.0 & 1.0\\
& G1 & 4.6M & 0.6 & 21.0 & 0.89 & 105 & 0.99 & 0.94 & 0.95 & 0.72\\
& G2 & 4.1M & 0.28 & 13.0 & 0.56 & 132 & 1.0 & 0.93 & 0.94 & 0.65\\
& G3 & 4.1M & 0.06 & 14.0 & 0.52 & 400 & 1.0 & 0.91 & 0.95 & 0.08\\
\hline
\end{tabular}

	}
	\caption{Multiscale results for biological (BIO) networks}\label{ms_bio}
\end{table}

\begin{table}[H]
	\centering
	\footnotesize
	\resizebox{\textwidth}{!}{%
		\begin{tabular}{ |l|l|l|l|l|l|l|l|l|l|l| }
		\hline
		\multirow{2}{*}{\textbf{Graph Name}} & \multirow{2}{*}{\textbf{Level}} & \multirow{2}{*}{\textbf{$|E|$}} & \multicolumn{8}{c|}{ \textbf{Properties} } \\
		\cline{4 - 11}
		& & & CC & D & Q & $\Gamma$ & $BC_{\rho}$ & $PR_{\rho}$ & $DC_{\rho}$ & $CC_{\rho}$\\
		\hline\multirow{4}{*}{ca-cit-Hepth}& G0 & 2.4M & 0.61 & 9.0 & 0.41 & 74 & 1.0 & 1.0 & 1.0 & 1.0\\
& G1 & 487.5K & 0.7 & 18.0 & 0.93 & 86 & 0.99 & 0.84 & 0.84 & 0.75\\
& G2 & 875.3K & 0.49 & 13.0 & 0.37 & 111 & 0.99 & 0.91 & 0.94 & 0.81\\
& G3 & 722.2K & 0.21 & 16.0 & 0.25 & 279 & 0.99 & 0.83 & 0.95 & -0.04\\
\hline
\multirow{4}{*}{cit-patent}& G0 & 16.5M & 0.09 & 26.0 & 0.81 & 3.6K & 1.0 & 1.0 & 1.0 & 1.0\\
& G1 & 5.8M & 0.16 & 61.0 & 0.93 & 43.8K & 1.0 & 0.79 & 0.78 & 0.73\\
& G2 & 3.4M & 0.13 & 57.0 & 0.97 & 631.4K & 1.0 & 0.58 & 0.64 & 0.59\\
& G3 & 5.0M & 0.01 & 39.0 & 0.84 & 235.5K & 1.0 & 0.68 & 0.74 & 0.54\\
\hline
\multirow{4}{*}{codblp}& G0 & 15.2M & 0.82 & 23.0 & 0.84 & 1 & 1.0 & 1.0 & 1.0 & 1.0\\
& G1 & 5.3M & 0.91 & 30.0 & 0.98 & 20.1K & 1.0 & 0.55 & 0.78 & 0.68\\
& G2 & 3.4M & 0.81 & 32.0 & 0.97 & 44.3K & 1.0 & 0.27 & 0.63 & 0.62\\
& G3 & 4.7M & 0.5 & 29.0 & 0.84 & 29.1K & 1.0 & 0.51 & 0.82 & 0.38\\
\hline
\end{tabular}

	}
	\caption{Multiscale results for citation (CIT) networks}\label{ms_cit}
\end{table}

\subsection{Running time}
\label{sub:runtimes}
\begin{figure}[H]
\centering
\begin{tabular}{cc}
	\includegraphics[width=0.45\linewidth, page=1]{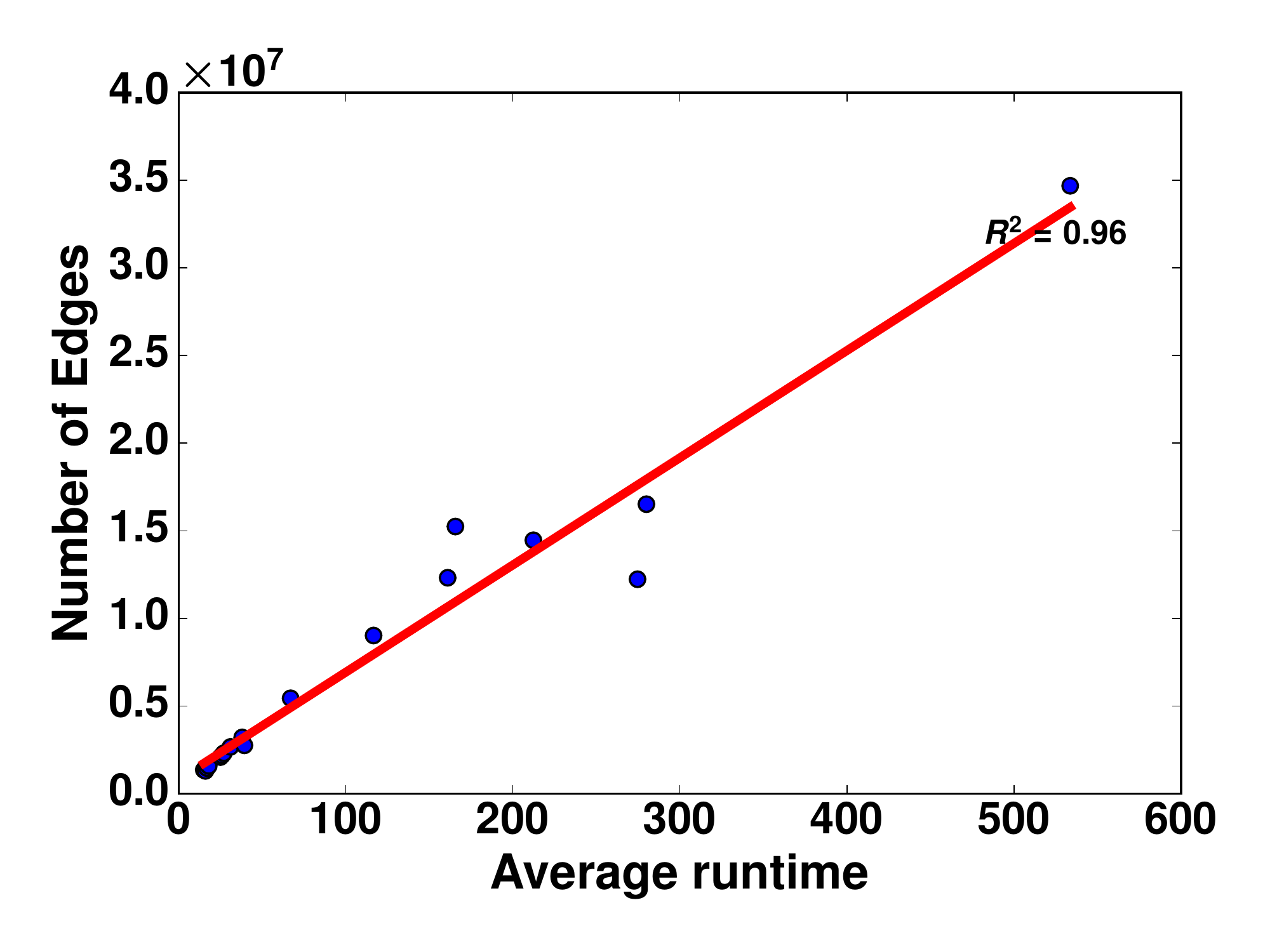}
&
	\includegraphics[width=0.45\linewidth, page=1]{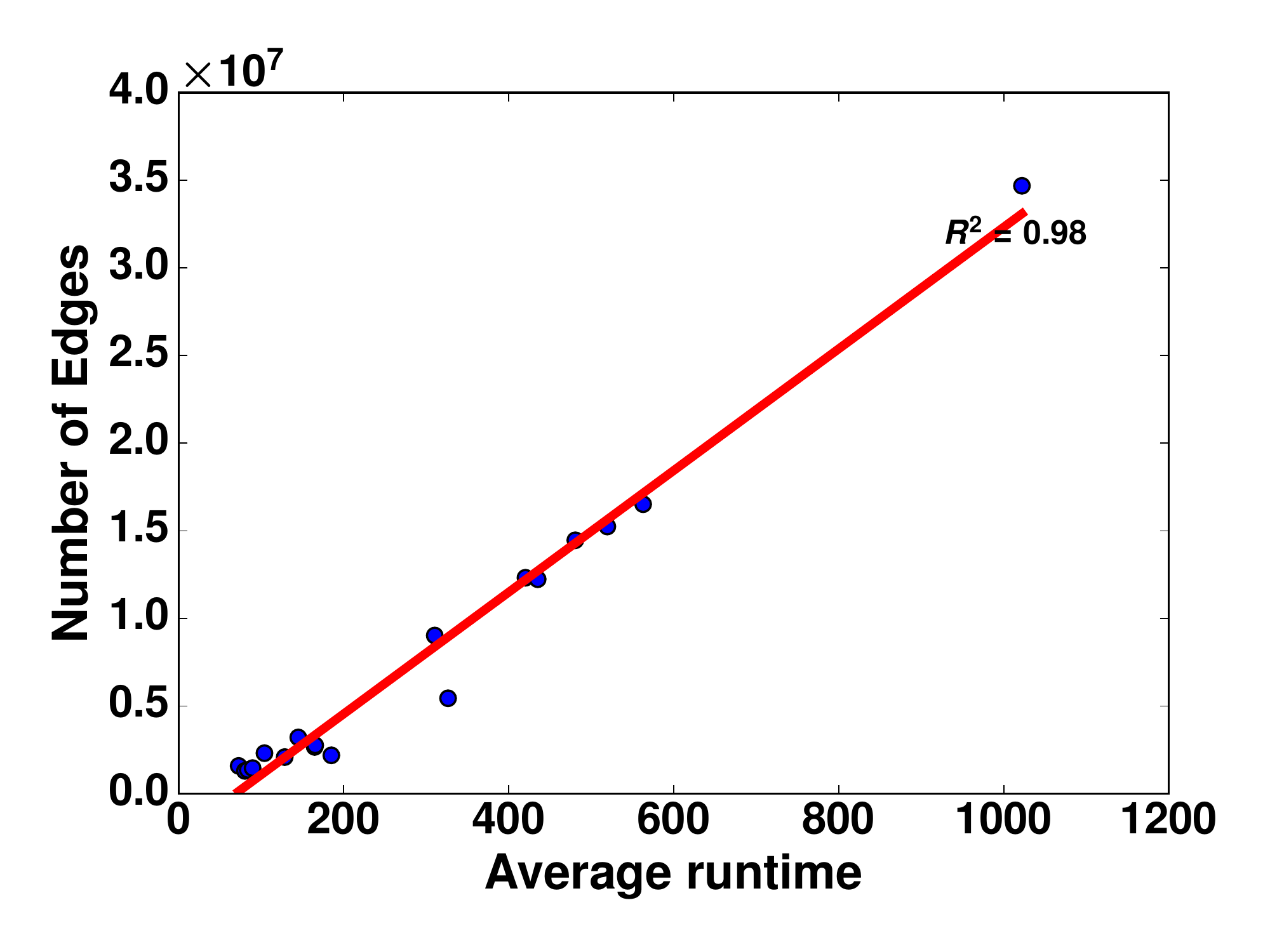}\\
	a) Single-level & b) Multi-level
\end{tabular}
		\caption[runtime-single.]{Running time of sparsification.}
	\label{runtime-sl}

\end{figure}

Figures \ref{runtime-sl}(a-b) show the running time of both single- and multi-level algorithms for varying sparsification ratios. Each point in the plot represents the number of edges in the graph versus the runtime in seconds averaged over three runs. The coefficient of determination, $R^{2}$, shows how well the regression line fits the model. An $R^2$ of 0 indicates the line does not fit the data and an $R^2$ of 1 indicates the line perfectly fits the data. The results show that both algorithms scales linearly with the number of edges in the graph. As mentioned earlier, this is important as it defeats the purpose of sparsification if the algorithm is slow. The experiments were performed in a Linux environment on a multicore compute server with 64 Intel Xeon cores and 64 GB of memory.

\subsection{Parallelization}
\begin{figure}[H]
	\centering
	\begin{tabular}{cc}
		\includegraphics[width=0.3\linewidth, page=1]{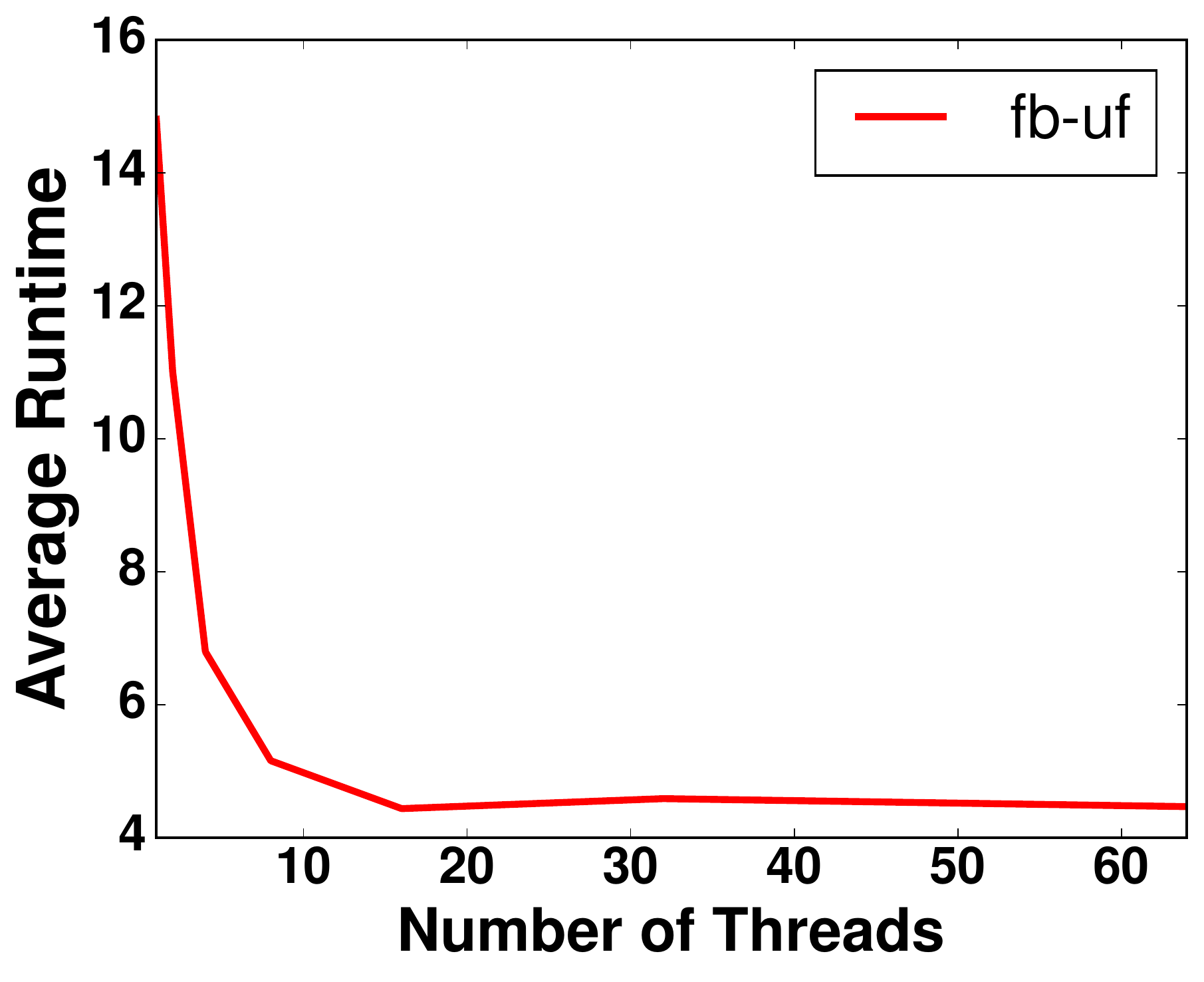}
		&
		\includegraphics[width=0.3\linewidth, page=2]{results/runtime-parallel-sep.pdf}\\
		\includegraphics[width=0.3\linewidth, page=3]{results/runtime-parallel-sep.pdf}
		&
		\includegraphics[width=0.3\linewidth, page=4]{results/runtime-parallel-sep.pdf}\\
	\end{tabular}
	\caption[runtime-parallel.]{Running time in shared memory model.}
	\label{runtime-par}
	
\end{figure}

Parallelization of the single-level algorithm does not require redesigning it. There are two computationally intensive parts of our method that gain from parallelization. One is the computation of the algebraic distance and the other the deletion of edges. Because of the implicitly parallel nature of the Jacobi over-relaxation, we are able to parallelize it by using OpenMP's shared data, multiple thread model. Since vector updates are independent, this method is highly efficient, creating speed gains of more than 50\% with only 8 threads as seen in Figure \ref{runtime-par}. Figure \ref{runtime-par} shows the benchmark results of parallelizing the algebraic distance computation where y-axis represents the average runtime averaged over 3 runs and x-axis represents the number of threads. We tested with number of threads ranging from 1 to 64 on 4 networks, namely, fb-uf, human-gene1, cit-patent, and catster. The experiments were performed in a Linux environment on a multicore compute server with 64 Intel Xeon cores and 64 GB of memory.

%
\section{Conclusions}
In this study we introduced single- and multi-level methods of network sparsification by algebraic distance. While many sparsification methods exist, most of them target certain properties without distinguishing short- and long-range connections that is the main goal of our method. We showed that by enabling different filtering capabilities, sparsification can be tuned to preserve either global or local structure or a combination of both.
In addition to preserving a host of graph properties, we believe that the development of the multilevel sparsification framework  can serve as a foundation for future work in that direction in which a variety of sparsification criteria (such as the algebraic distance) can be incorporated into it.

\appendix
\section{Normalized Sparsification} \label{App:AppendixA}

\begin{figure}[H]
\footnotesize
\hspace{1cm}
\begin{tabular}{m{0cm} m{1cm} m{3.6cm} m{3.6cm} m{3.6cm}}
	\centering
\multirow{8}{*}{\hspace{-2cm}\includegraphics[angle=90,page=25]{results/plots/normalized/fb-indiana_fb-uf_fb-texas84_fb-penn94_gplus_lj.pdf}} & &
\centering $\delta$-weak & \centering Mixture & \centering $\delta$-strong \arraybackslash \\
& \raisebox{.25\height}{\rotatebox{90}{\parbox{1.9cm}{Betweenness\\Centrality}}} &
\includegraphics[width=\linewidth, page=1]{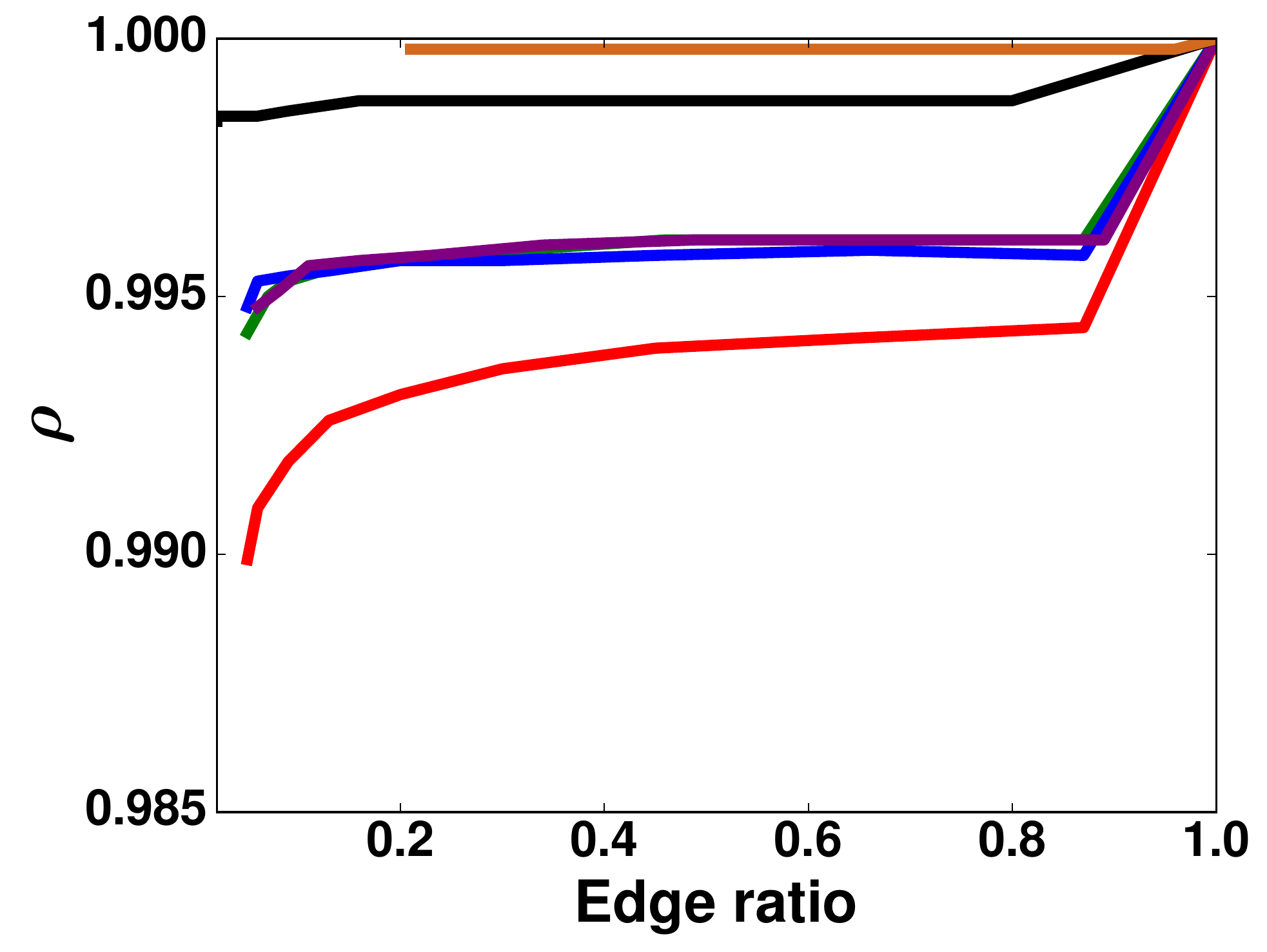} &
\includegraphics[width=\linewidth, page=2]{results/plots/normalized/fb-indiana_fb-uf_fb-texas84_fb-penn94_gplus_lj.pdf} &
\includegraphics[width=\linewidth, page=3]{results/plots/normalized/fb-indiana_fb-uf_fb-texas84_fb-penn94_gplus_lj.pdf}\\
& \raisebox{.25\height}{\rotatebox{90}{\parbox{1.9cm}{Clustering\\Coefficient\\Distribution}}} &
\includegraphics[width=\linewidth, page=7]{results/plots/normalized/fb-indiana_fb-uf_fb-texas84_fb-penn94_gplus_lj.pdf} &
\includegraphics[width=\linewidth, page=8]{results/plots/normalized/fb-indiana_fb-uf_fb-texas84_fb-penn94_gplus_lj.pdf} &
\includegraphics[width=\linewidth, page=9]{results/plots/normalized/fb-indiana_fb-uf_fb-texas84_fb-penn94_gplus_lj.pdf}\\
& \raisebox{.25\height}{\rotatebox{90}{\parbox{1.9cm}{Degree\\Centrality}}} &
\includegraphics[width=\linewidth, page=10]{results/plots/normalized/fb-indiana_fb-uf_fb-texas84_fb-penn94_gplus_lj.pdf} &
\includegraphics[width=\linewidth, page=11]{results/plots/normalized/fb-indiana_fb-uf_fb-texas84_fb-penn94_gplus_lj.pdf} &
\includegraphics[width=\linewidth, page=12]{results/plots/normalized/fb-indiana_fb-uf_fb-texas84_fb-penn94_gplus_lj.pdf}\\
& \raisebox{.25\height}{\rotatebox{90}{\parbox{1.9cm}{Diameter}}} &
\includegraphics[width=\linewidth, page=13]{results/plots/normalized/fb-indiana_fb-uf_fb-texas84_fb-penn94_gplus_lj.pdf} &
\includegraphics[width=\linewidth, page=14]{results/plots/normalized/fb-indiana_fb-uf_fb-texas84_fb-penn94_gplus_lj.pdf} &
\includegraphics[width=\linewidth, page=15]{results/plots/normalized/fb-indiana_fb-uf_fb-texas84_fb-penn94_gplus_lj.pdf}\\
& \raisebox{.25\height}{\rotatebox{90}{\parbox{1.9cm}{Components}}} &
\includegraphics[width=\linewidth, page=16]{results/plots/normalized/fb-indiana_fb-uf_fb-texas84_fb-penn94_gplus_lj.pdf} &
\includegraphics[width=\linewidth, page=17]{results/plots/normalized/fb-indiana_fb-uf_fb-texas84_fb-penn94_gplus_lj.pdf} &
\includegraphics[width=\linewidth, page=18]{results/plots/normalized/fb-indiana_fb-uf_fb-texas84_fb-penn94_gplus_lj.pdf}\\
& \raisebox{.25\height}{\rotatebox{90}{\parbox{1.9cm}{Modularity}}} &
\includegraphics[width=\linewidth, page=19]{results/plots/normalized/fb-indiana_fb-uf_fb-texas84_fb-penn94_gplus_lj.pdf} &
\includegraphics[width=\linewidth, page=20]{results/plots/normalized/fb-indiana_fb-uf_fb-texas84_fb-penn94_gplus_lj.pdf} &
\includegraphics[width=\linewidth, page=21]{results/plots/normalized/fb-indiana_fb-uf_fb-texas84_fb-penn94_gplus_lj.pdf}\\
& \raisebox{.25\height}{\rotatebox{90}{\parbox{1.9cm}{Pagerank}}} &
\includegraphics[width=\linewidth, page=22]{results/plots/normalized/fb-indiana_fb-uf_fb-texas84_fb-penn94_gplus_lj.pdf} &
\includegraphics[width=\linewidth, page=23]{results/plots/normalized/fb-indiana_fb-uf_fb-texas84_fb-penn94_gplus_lj.pdf} &
\includegraphics[width=\linewidth, page=24]{results/plots/normalized/fb-indiana_fb-uf_fb-texas84_fb-penn94_gplus_lj.pdf}\\
\end{tabular}
\caption{Social Networks 1}\label{fig:norm_soc1}
\end{figure}

\begin{figure}[H]
\footnotesize
\hspace{1cm}
\begin{tabular}{m{0cm} m{1cm} m{3.6cm} m{3.6cm} m{3.6cm}}
	\centering
\multirow{8}{*}{\hspace{-2cm}\includegraphics[angle=90,page=25]{results/plots/normalized/flickr_foursquare_catster_buzznet_blogcatalog_livemocha.pdf}} & &
\centering $\delta$-weak & \centering Mixture & \centering $\delta$-strong \arraybackslash \\
& \raisebox{.25\height}{\rotatebox{90}{\parbox{1.9cm}{Betweenness\\Centrality}}} &
\includegraphics[width=\linewidth, page=1]{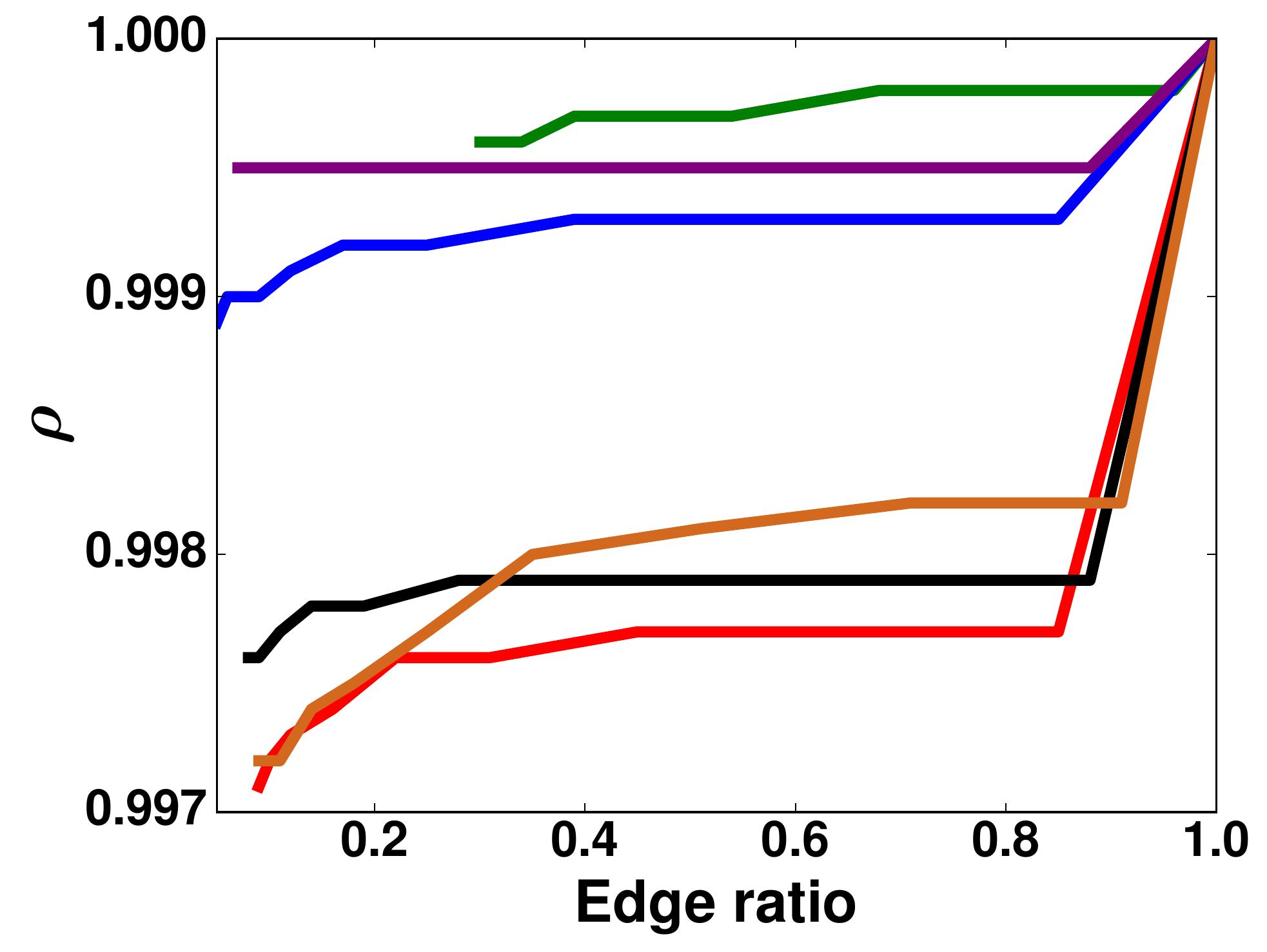} &
\includegraphics[width=\linewidth, page=2]{results/plots/normalized/flickr_foursquare_catster_buzznet_blogcatalog_livemocha.pdf} &
\includegraphics[width=\linewidth, page=3]{results/plots/normalized/flickr_foursquare_catster_buzznet_blogcatalog_livemocha.pdf}\\
& \raisebox{.25\height}{\rotatebox{90}{\parbox{1.9cm}{Clustering\\Coefficient\\Distribution}}} &
\includegraphics[width=\linewidth, page=7]{results/plots/normalized/flickr_foursquare_catster_buzznet_blogcatalog_livemocha.pdf} &
\includegraphics[width=\linewidth, page=8]{results/plots/normalized/flickr_foursquare_catster_buzznet_blogcatalog_livemocha.pdf} &
\includegraphics[width=\linewidth, page=9]{results/plots/normalized/flickr_foursquare_catster_buzznet_blogcatalog_livemocha.pdf}\\
& \raisebox{.25\height}{\rotatebox{90}{\parbox{1.9cm}{Degree\\Centrality}}} &
\includegraphics[width=\linewidth, page=10]{results/plots/normalized/flickr_foursquare_catster_buzznet_blogcatalog_livemocha.pdf} &
\includegraphics[width=\linewidth, page=11]{results/plots/normalized/flickr_foursquare_catster_buzznet_blogcatalog_livemocha.pdf} &
\includegraphics[width=\linewidth, page=12]{results/plots/normalized/flickr_foursquare_catster_buzznet_blogcatalog_livemocha.pdf}\\
& \raisebox{.25\height}{\rotatebox{90}{\parbox{1.9cm}{Diameter}}} &
\includegraphics[width=\linewidth, page=13]{results/plots/normalized/flickr_foursquare_catster_buzznet_blogcatalog_livemocha.pdf} &
\includegraphics[width=\linewidth, page=14]{results/plots/normalized/flickr_foursquare_catster_buzznet_blogcatalog_livemocha.pdf} &
\includegraphics[width=\linewidth, page=15]{results/plots/normalized/flickr_foursquare_catster_buzznet_blogcatalog_livemocha.pdf}\\
& \raisebox{.25\height}{\rotatebox{90}{\parbox{1.9cm}{Components}}} &
\includegraphics[width=\linewidth, page=16]{results/plots/normalized/flickr_foursquare_catster_buzznet_blogcatalog_livemocha.pdf} &
\includegraphics[width=\linewidth, page=17]{results/plots/normalized/flickr_foursquare_catster_buzznet_blogcatalog_livemocha.pdf} &
\includegraphics[width=\linewidth, page=18]{results/plots/normalized/flickr_foursquare_catster_buzznet_blogcatalog_livemocha.pdf}\\
& \raisebox{.25\height}{\rotatebox{90}{\parbox{1.9cm}{Modularity}}} &
\includegraphics[width=\linewidth, page=19]{results/plots/normalized/flickr_foursquare_catster_buzznet_blogcatalog_livemocha.pdf} &
\includegraphics[width=\linewidth, page=20]{results/plots/normalized/flickr_foursquare_catster_buzznet_blogcatalog_livemocha.pdf} &
\includegraphics[width=\linewidth, page=21]{results/plots/normalized/flickr_foursquare_catster_buzznet_blogcatalog_livemocha.pdf}\\
& \raisebox{.25\height}{\rotatebox{90}{\parbox{1.9cm}{Pagerank}}} &
\includegraphics[width=\linewidth, page=22]{results/plots/normalized/flickr_foursquare_catster_buzznet_blogcatalog_livemocha.pdf} &
\includegraphics[width=\linewidth, page=23]{results/plots/normalized/flickr_foursquare_catster_buzznet_blogcatalog_livemocha.pdf} &
\includegraphics[width=\linewidth, page=24]{results/plots/normalized/flickr_foursquare_catster_buzznet_blogcatalog_livemocha.pdf}\\
\end{tabular}
\caption{Social Networks 2}\label{fig:norm_soc2}
\end{figure}

\begin{figure}[H]
\footnotesize
\hspace{1cm}
\begin{tabular}{m{0cm} m{1cm} m{3.6cm} m{3.6cm} m{3.6cm}}
	\centering
\multirow{8}{*}{\hspace{-2cm}\includegraphics[angle=90,page=25]{results/plots/normalized/ca-cit-Hepth_codblp_cit-patent.pdf}} & &
\centering $\delta$-weak & \centering Mixture & \centering $\delta$-strong \arraybackslash \\
& \raisebox{.25\height}{\rotatebox{90}{\parbox{1.9cm}{Betweenness\\Centrality}}} &
\includegraphics[width=\linewidth, page=1]{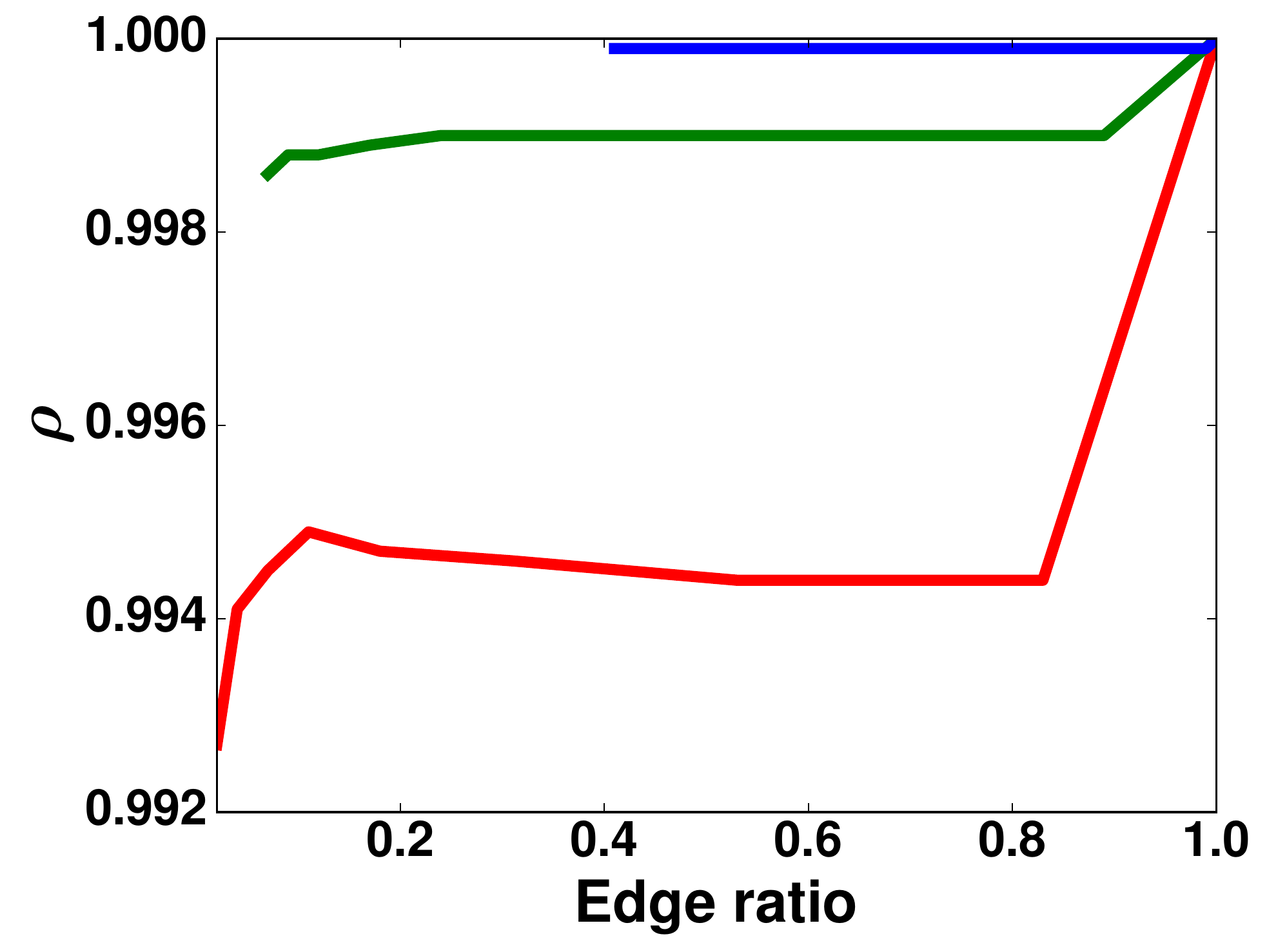} &
\includegraphics[width=\linewidth, page=2]{results/plots/normalized/ca-cit-Hepth_codblp_cit-patent.pdf} &
\includegraphics[width=\linewidth, page=3]{results/plots/normalized/ca-cit-Hepth_codblp_cit-patent.pdf}\\
& \raisebox{.25\height}{\rotatebox{90}{\parbox{1.9cm}{Clustering\\Coefficient\\Distribution}}} &
\includegraphics[width=\linewidth, page=7]{results/plots/normalized/ca-cit-Hepth_codblp_cit-patent.pdf} &
\includegraphics[width=\linewidth, page=8]{results/plots/normalized/ca-cit-Hepth_codblp_cit-patent.pdf} &
\includegraphics[width=\linewidth, page=9]{results/plots/normalized/ca-cit-Hepth_codblp_cit-patent.pdf}\\
& \raisebox{.25\height}{\rotatebox{90}{\parbox{1.9cm}{Degree\\Centrality}}} &
\includegraphics[width=\linewidth, page=10]{results/plots/normalized/ca-cit-Hepth_codblp_cit-patent.pdf} &
\includegraphics[width=\linewidth, page=11]{results/plots/normalized/ca-cit-Hepth_codblp_cit-patent.pdf} &
\includegraphics[width=\linewidth, page=12]{results/plots/normalized/ca-cit-Hepth_codblp_cit-patent.pdf}\\
& \raisebox{.25\height}{\rotatebox{90}{\parbox{1.9cm}{Diameter}}} &
\includegraphics[width=\linewidth, page=13]{results/plots/normalized/ca-cit-Hepth_codblp_cit-patent.pdf} &
\includegraphics[width=\linewidth, page=14]{results/plots/normalized/ca-cit-Hepth_codblp_cit-patent.pdf} &
\includegraphics[width=\linewidth, page=15]{results/plots/normalized/ca-cit-Hepth_codblp_cit-patent.pdf}\\
& \raisebox{.25\height}{\rotatebox{90}{\parbox{1.9cm}{Components}}} &
\includegraphics[width=\linewidth, page=16]{results/plots/normalized/ca-cit-Hepth_codblp_cit-patent.pdf} &
\includegraphics[width=\linewidth, page=17]{results/plots/normalized/ca-cit-Hepth_codblp_cit-patent.pdf} &
\includegraphics[width=\linewidth, page=18]{results/plots/normalized/ca-cit-Hepth_codblp_cit-patent.pdf}\\
& \raisebox{.25\height}{\rotatebox{90}{\parbox{1.9cm}{Modularity}}} &
\includegraphics[width=\linewidth, page=19]{results/plots/normalized/ca-cit-Hepth_codblp_cit-patent.pdf} &
\includegraphics[width=\linewidth, page=20]{results/plots/normalized/ca-cit-Hepth_codblp_cit-patent.pdf} &
\includegraphics[width=\linewidth, page=21]{results/plots/normalized/ca-cit-Hepth_codblp_cit-patent.pdf}\\
& \raisebox{.25\height}{\rotatebox{90}{\parbox{1.9cm}{Pagerank}}} &
\includegraphics[width=\linewidth, page=22]{results/plots/normalized/ca-cit-Hepth_codblp_cit-patent.pdf} &
\includegraphics[width=\linewidth, page=23]{results/plots/normalized/ca-cit-Hepth_codblp_cit-patent.pdf} &
\includegraphics[width=\linewidth, page=24]{results/plots/normalized/ca-cit-Hepth_codblp_cit-patent.pdf}\\
\end{tabular}
\caption{Citation Networks}\label{fig:norm_cit}
\end{figure}

\begin{figure}[H]
\footnotesize
\hspace{1cm}
\begin{tabular}{m{0cm} m{1cm} m{3.6cm} m{3.6cm} m{3.6cm}}
	\centering
\multirow{8}{*}{\hspace{-2cm}\includegraphics[angle=90,page=25]{results/plots/normalized/human1_human2_mouse.pdf}} & &
\centering $\delta$-weak & \centering Mixture & \centering $\delta$-strong \arraybackslash \\
& \raisebox{.25\height}{\rotatebox{90}{\parbox{1.9cm}{Betweenness\\Centrality}}} &
\includegraphics[width=\linewidth, page=1]{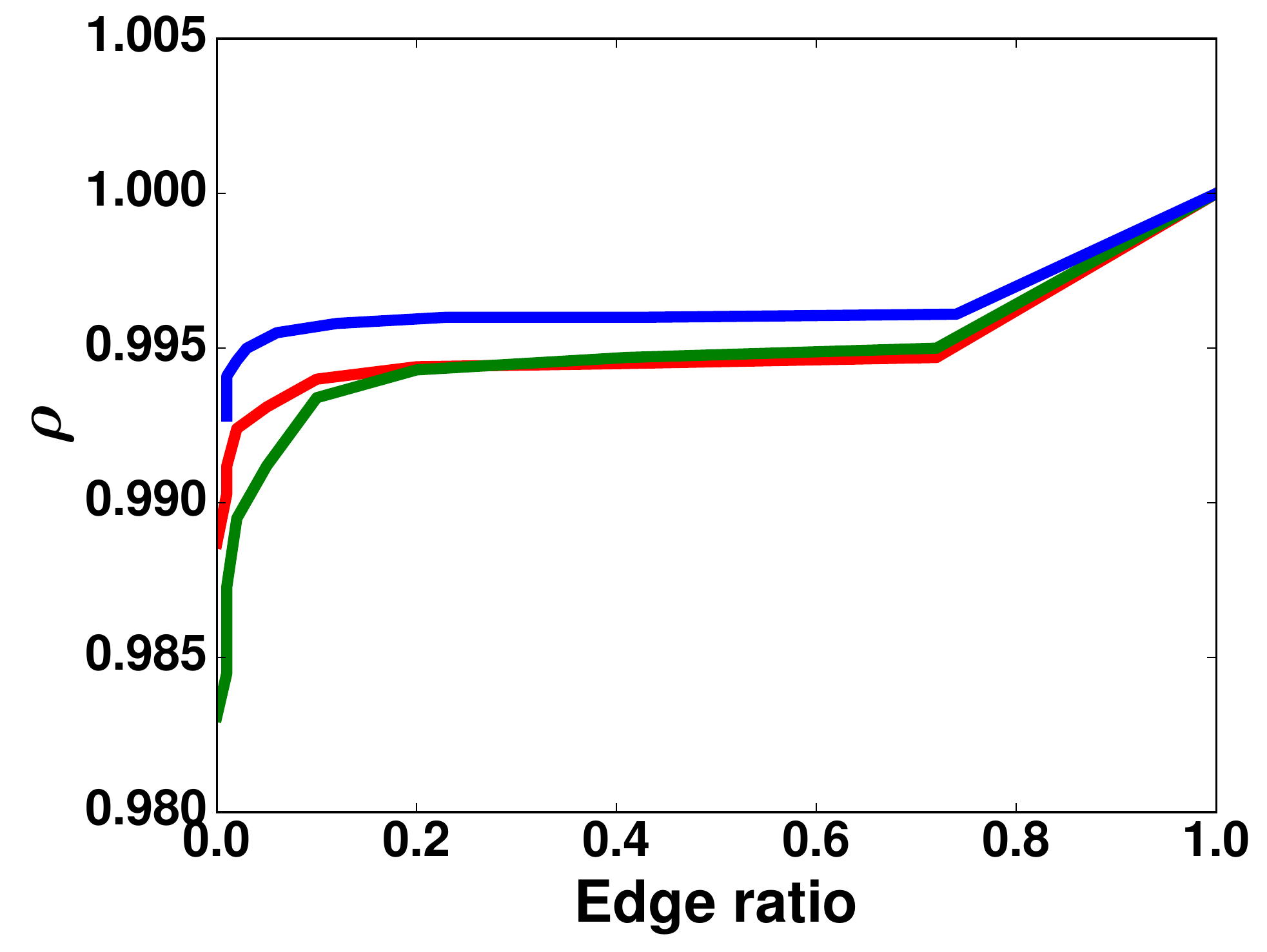} &
\includegraphics[width=\linewidth, page=2]{results/plots/normalized/human1_human2_mouse.pdf} &
\includegraphics[width=\linewidth, page=3]{results/plots/normalized/human1_human2_mouse.pdf}\\
& \raisebox{.25\height}{\rotatebox{90}{\parbox{1.9cm}{Clustering\\Coefficient\\Distribution}}} &
\includegraphics[width=\linewidth, page=7]{results/plots/normalized/human1_human2_mouse.pdf} &
\includegraphics[width=\linewidth, page=8]{results/plots/normalized/human1_human2_mouse.pdf} &
\includegraphics[width=\linewidth, page=9]{results/plots/normalized/human1_human2_mouse.pdf}\\
& \raisebox{.25\height}{\rotatebox{90}{\parbox{1.9cm}{Degree\\Centrality}}} &
\includegraphics[width=\linewidth, page=10]{results/plots/normalized/human1_human2_mouse.pdf} &
\includegraphics[width=\linewidth, page=11]{results/plots/normalized/human1_human2_mouse.pdf} &
\includegraphics[width=\linewidth, page=12]{results/plots/normalized/human1_human2_mouse.pdf}\\
& \raisebox{.25\height}{\rotatebox{90}{\parbox{1.9cm}{Diameter}}} &
\includegraphics[width=\linewidth, page=13]{results/plots/normalized/human1_human2_mouse.pdf} &
\includegraphics[width=\linewidth, page=14]{results/plots/normalized/human1_human2_mouse.pdf} &
\includegraphics[width=\linewidth, page=15]{results/plots/normalized/human1_human2_mouse.pdf}\\
& \raisebox{.25\height}{\rotatebox{90}{\parbox{1.9cm}{Components}}} &
\includegraphics[width=\linewidth, page=16]{results/plots/normalized/human1_human2_mouse.pdf} &
\includegraphics[width=\linewidth, page=17]{results/plots/normalized/human1_human2_mouse.pdf} &
\includegraphics[width=\linewidth, page=18]{results/plots/normalized/human1_human2_mouse.pdf}\\
& \raisebox{.25\height}{\rotatebox{90}{\parbox{1.9cm}{Modularity}}} &
\includegraphics[width=\linewidth, page=19]{results/plots/normalized/human1_human2_mouse.pdf} &
\includegraphics[width=\linewidth, page=20]{results/plots/normalized/human1_human2_mouse.pdf} &
\includegraphics[width=\linewidth, page=21]{results/plots/normalized/human1_human2_mouse.pdf}\\
& \raisebox{.25\height}{\rotatebox{90}{\parbox{1.9cm}{Pagerank}}} &
\includegraphics[width=\linewidth, page=22]{results/plots/normalized/human1_human2_mouse.pdf} &
\includegraphics[width=\linewidth, page=23]{results/plots/normalized/human1_human2_mouse.pdf} &
\includegraphics[width=\linewidth, page=24]{results/plots/normalized/human1_human2_mouse.pdf}\\
\end{tabular}
\caption{Biological Networks}\label{fig:norm_bio}
\end{figure}

We experimented with the single-level algorithm that employ  normalized algebraic distances (see line 15 of Algorithm \ref{alg1}). The purpose of this normalization is to decrease the strength of connection expressed in the algebraic distance between hubs. The normalization results show that normalizing the algebraic distance further improves properties that are sensitive to the existence of weak edges. Example are diameter and connected components. As seen in the plots for diameter (see $\delta$-weak column in Figures \ref{fig:norm_soc1}, \ref{fig:norm_soc2}, \ref{fig:norm_cit}, and \ref{fig:norm_bio}), the minimum edge ratio before the diameter deteriorates is further improved. Similarly for the number of components the number of components for the smallest sparse graph is reduced and some case kept constant as seen in $\delta$-weak column in Figures \ref{fig:norm_soc1}, \ref{fig:norm_soc2}, \ref{fig:norm_cit}, and \ref{fig:norm_bio}). Such properties as local clustering coefficient, degree centrality, and PageRank that do not depend on global edges are relatively unaffected.


%

\footnotesize
\normalsize

\footnotesize
\normalsize

\footnotesize
\normalsize

\footnotesize
\normalsize

\bibliographystyle{plain}
\bibliography{fullbib,ej-bibliography}

\end{document}